\documentclass[aps, prb, twocolumn, showpacs, superscriptaddress, preprintnumbers, bibnotes, amsmath, amssymb]{revtex4-1}

\usepackage{graphicx}
\usepackage{dcolumn}
\usepackage{bm}
\usepackage[flushleft]{threeparttable}
\usepackage{autobreak}
\usepackage{bbold}
\usepackage{amsmath}
\usepackage{amssymb}
\usepackage{braket}
\usepackage{multirow}
\usepackage{array}
\usepackage{url}
\usepackage{booktabs}
\usepackage{float}
\usepackage{hyperref}
\usepackage{mathtools}

\newcolumntype{P}[1]{>{\centering\arraybackslash}p{#1}}
\usepackage[all]{hypcap}

\usepackage{cases}

\newcommand{\bk}{{\bf k}}
\newcommand{\bK}{{\bf K}}
\newcommand{\bq}{{\bf q}}

\newcommand{\bb}{{\bf b}}

\newcommand{\bd}{{\bf d}}

\newcommand{\bn}{{\bf n}}

\newcommand{\up}{\uparrow}
\newcommand{\down}{\downarrow}

\newcommand{\bsigma}{\boldsymbol{\sigma} }

\newcommand{\be}{\begin{equation}}
\newcommand{\ee}{\end{equation}}
\newcommand{\beq}{\begin{eqnarray}}
\newcommand{\eeq}{\end{eqnarray}}

\newcommand{\makebf}[1]{\boldsymbol{#1} }

\usepackage{color}
\hypersetup{colorlinks=true, linkcolor=blue,urlcolor=blue, citecolor=cyan, pdfborder={0 0 0},}

\begin{document}
\title{Noncentrosymmetric topological Dirac semimetals in three dimensions}

\author{Heng Gao}
\affiliation{Songshan Lake Materials Laboratory, Dongguan, Guangdong 523808, China}
\affiliation{Beijing National Laboratory for Condensed Matter Physics, Institute of Physics, Chinese Academy of Sciences, Beijing 100190, China}

\author{Jeremy Strockoz}
\affiliation{Department of Physics, Drexel University, Philadelphia, PA 19104, USA}%

\author{Mario Frakulla}
\affiliation{Department of Physics, Drexel University, Philadelphia, PA 19104, USA}%
\affiliation{Department of Mechanical Engineering, Drexel University, Philadelphia, PA 19104, USA}%

\author{J\"orn W. F. Venderbos}
\email{jwv34@drexel.edu}
\affiliation{Department of Physics, Drexel University, Philadelphia, PA 19104, USA}%
\affiliation{Department of Materials Science \& Engineering, Drexel University, Philadelphia, PA 19104, USA}%

\author{Hongming Weng}%
\email{hmweng@iphy.ac.cn}
\affiliation{Beijing National Laboratory for Condensed Matter Physics, Institute of Physics, Chinese Academy of Sciences, Beijing 100190, China}
\affiliation{Songshan Lake Materials Laboratory, Dongguan, Guangdong 523808, China}

\date{\today}

\begin{abstract}
Topological Dirac semimetals are a class of semimetals that host symmetry-protected Dirac points near the Fermi level, which arise due to a band inversion of the conduction and valence bands. In this work, we study the less explored class of \emph{noncentrosymmetric} topological Dirac semimetals in three dimensions. We identify the noncentrosymmetric crystallographic point groups required to stabilize fourfold degenerate band crossings and derive model Hamiltonians for all distinct types of band inversions allowed by symmetry. Using these model Hamiltonians, which emphasize the physical nature of the allowed couplings, we establish the generic electronic phase diagram noncentrosymmetric Dirac semimetals and show that it generically includes phases with coexistent Weyl point nodes or Weyl line nodes. In particular, for one specific type of band inversion in sixfold symmetric systems we show that Weyl line nodes are always present. Based on first-principles calculations, we predict that BiPd$_2$O$_4$ is a noncentrosymmetric Dirac semimetal under 20 Gpa pressure and hosts topological type-II Dirac points on the fourfold rotation axis. Furthermore, we propose that the hexagonal polar alloy LiZnSb$_{x}$Bi$_{1-x}$ realizes a Dirac semimetal with coexistent Weyl points. Interestingly, the emergence and location of the Weyl points is highly tunable and can be controlled by the alloy concentration $x$. More generally, our results not only establish band-inverted noncentrosymmetric systems as a broad and versatile class of topological semimetals, but also provide a framework for studying the quantum nonlinear Hall effect and nonlinear optical properties in the Dirac semimetals. \par
\end{abstract}

\maketitle

\section{Introduction}

Symmetry and topology are two important concepts from mathematics which have found increasingly many important applications in condensed mater physics in recent years.\cite{Hasan10p3045,Armitage18p015001} Both crystalline and time-reversal symmetries not only determine macroscopic properties of solids but also powerfully affect the topological electronic properties of matter. The interplay between (crystalline) symmetry and the topology of electronic structure in solids leads to a variety of new topological states, which include most notably topological insulators \cite{Kane05p226801,Hasan10p3045,Weng14p849}, topological crystalline insulators \cite{Fu11p106802,Song19peaax2007}, and topological semimetals \cite{Armitage18p015001,Gao19p153,Weng16p303001,Bernevig18p041001,Weng17p798}. Topological semimetals are gapless systems hosting topologically protected band crossings near the Fermi level, which can be categorized based on codimension (e.g. point nodes, line nodes) and degeneracy of the crossing (e.g. twofold, fourfold, etc.). This categorization gives rise to different families of topological semimetals such as Dirac semimetals \cite{Yong12p14045,Wang12p195320}, Weyl semimetals \cite{Wan11p205101,Weng15p011029,Lv15p031013,Liu19p1282}, double Dirac semimetals \cite{Wieder16p186402}, nodal line semimetals \cite{Weng13p045108,Kim15p036806,Yu15p036807,Fang16p117106}, nodal surface semimetals \cite{Liang16p085427,Wu18p115125} and multifold fermion semimetals \cite{Bradlyn16p5037,Lv17627}. \par

Dirac semimetals are defined by the presence of stable fourfold degenerate Dirac fermions in the momentum space and are an important member of the class of topological semimetals.  
Two types of Dirac semimetals can be distinguished: symmetry-enforced Dirac semimetals and topological Dirac semimetals~\cite{Yang14p1}. This distinction relies on the symmetry protection mechanism. Whereas symmetry-enforced Dirac semimetals arise as a result of nonsymmorphic symmetry-mandated degeneracies at high symmetry points on the boundary of Brillouin zone (BZ), and are therefore pinned, topological Dirac semimetals arise as a result of a band inversion. In the band-inverted regime, rotation crystal symmetry can protect the inversion-induced crossings on the rotation axis, which are not pinned to a particular point on the rotation axis. \par

Young {\it et al.} were the first to show that nonsymmorphic symmetries can protect the Dirac points at the boundary of the BZ and proposed $\beta$-cristobalite SiO$_{2}$ as a possible material candidate \cite{Yong12p14045}. Unfortunately, $\beta$-cristobalite SiO$_{2}$ is weakly metastable in nature. Recently, however, a number of nonsymmorphic symmetry protected Dirac semimetals have been predicted in distorted spinels BiZnSiO$_{4}$, BiCaSiO$_{4}$, BiAlInO$_{4}$, and BiMgSiO$_{4}$ and molybdenum monochalcogenide compounds AI(MoXVI)$_{3}$ (AI= Na, K, Rb, In, Tl; XVI= S, Se, Te) \cite{Liu17p021019}, but to date the experimental realization and verification of symmetry-enforced Dirac semimetals remain an open problem. In contrast, realizations of topological Dirac semimetals have been demonstrated in photoemission experiments, such as in Na$_3$Bi \cite{Liu14p864} and Cd$_2$As$_3$ \cite{Liu14p677}---by now two well-known Dirac semimetal materials. These experimental discoveries followed theoretical predictions based on first-principles calculations~\cite{Wang12p195320,Wang13p125427}, highlighting the importance of first-principles electronic structure methods for topological materials search strategies. \par

The initial proposals for Dirac semimetals in three dimensions generated great excitement and inspired a collective effort to characterize their properties and find more material realizations of Dirac semimetals. In the case of systems with inversion symmetry this effort was facilitated by a general symmetry classification of centrosymmetric Dirac semimetals~\cite{Yang14p1}, which has led to the prediction of a large number of Dirac semimetals with inversion symmetry, including BaAgBi \cite{Gibson15p205128}, SrPd$_{3}$O$_{4}$ \cite{Li17p035102} and MgTa$_2$N$_3$ \cite{Wu18p081115}. Compared to the centrosymmetric variants, Dirac semimetals without inversion symmetry have been much less studied and theoretical predictions of material candidates are comparatively rare \cite{Gao16p205109, Cao17p115203, Chen17p044201,Xia20p041201}. Furthermore, a detailed understanding of the phase diagram of noncentrosymmetric Dirac semimetals, obtained from a derivation and analysis of effective model Hamiltonians, is still lacking, even though the absence of inversion symmetry is expected to give rise to a richer phase diagram. \par

In this work, we systematically study the possible realizations of noncentrosymmetric topological Dirac semimetals in three dimensional crystals. Based on a point group symmetry analysis, we derive low-energy $k\cdot p$-type model Hamiltonians for all distinct band-inversion induced Dirac semimetals without inversion symmetry, using a formulation which emphasizes the physical nature of the allowed couplings. Based on the obtained model Hamiltonians, we determine the generic electronic phase diagram of noncentrosymmetric band-inverted topological semimetals. We show that two types of Dirac points can exist: ({\it i}) conventional fourfold degenerate point nodal band crossings with linear dispersion away from the crossing and ({\it ii}) Dirac points at which Weyl nodal lines terminate. In particular, for some types of band inversions Weyl line nodes always exist. We furthermore show that the generic phase diagram includes a phase with coexistent Dirac and Weyl fermions. \par

We carry out first-principles calculations to predict two material candidates. In particular, we predict that a new noncentrosymmetric topological Dirac semimetal is realized in BiPd$_2$O$_4$ under 20 Gpa pressure, which has $C_{4v}$ point group symmetry and hosts type-II noncentrosymmetric Dirac Fermions. In addition, we propose that the alloy LiZnSb$_{x}$Bi$_{1-x}$ realizes a highly tunable topologically semimetallic phase with coexistent Dirac and Weyl points. The existence and the location of the Weyl points can be controlled by the alloy concentration of LiZnSb$_{x}$Bi$_{1-x}$, providing a tuning parameter directly accessible in experiment. \par

\section{Topological Dirac semimetals without inversion symmetry \label{sec:ham}}

\subsection{General symmetry analysis \label{ssec:symmetry}}

\begin{figure}[pt]
    \centering
    \includegraphics[width=8.6cm]{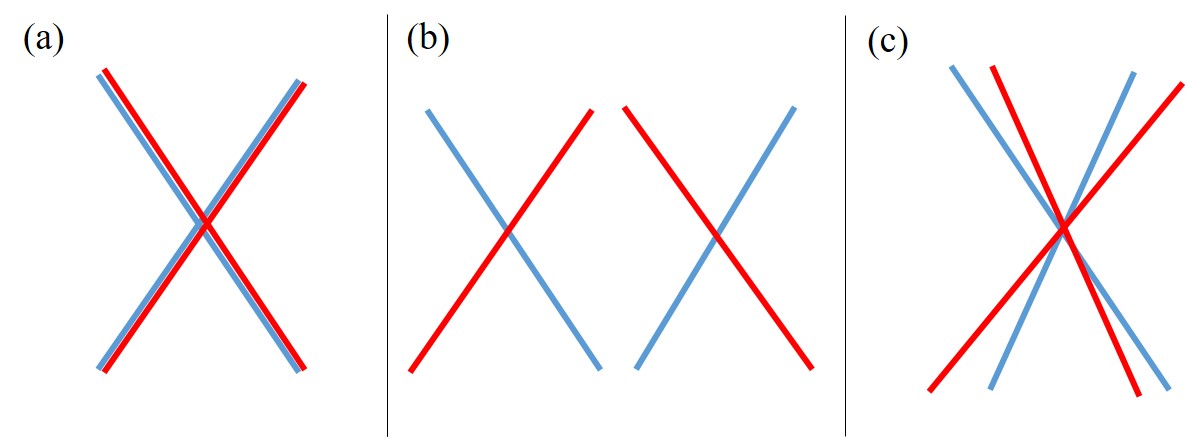}
   \caption{Schematic band structures of (a) centrosymmetric Dirac semimetals (b) Weyl semimetals (c) noncentrosymmetric Dirac semimetals. The red and blue colors represent difference of energy bands.}
    \label{fig:fig_1}
\end{figure}

We begin by reviewing the general symmetry requirements a (strongly) spin-orbit coupled material must satisfy in order to realize a noncentrosymmetric topological Dirac semimetal. Since we exclusively focus on band inversions at $\Gamma$ in this work, the relevant symmetry group is just the crystal point group. For a material to potentially realize a noncentrosymmetric Dirac semimetal, its point group must satisfy two main criteria: ({\it i}) First, by assumption, it cannot contain the inversion and ({\it ii}) second, the point group must contain a rotation axis. The presence of a rotation axis is important, since discrete crystal rotation symmetry can protect band degeneracies on the rotation axis in momentum space, provided the bands which cross have different rotation eigenvalues \cite{Fang12p266802, Yang14p1, Gao16p205109}. For such degeneracies to realize Dirac points (i.e., fourfold degenerate band crossings), it is further necessary that the double group of symmetries which leave the rotation axis invariant admits two-dimensional representations. A Dirac point then occurs when two bands characterized by different two-dimensional representations cross on the rotation axis. By an exhaustive search of all noncentrosymmetric point groups, we find that only the groups $C_{4v}$ and $C_{6v}$ satisfy these criteria, and we will therefore focus on these two groups in all that follows. \par

The basic difference between Dirac points in noncentrosymmetric systems and other types of topological semimetallic band crossings, such as centrosymmetric Dirac points and Weyl points, is schematically illustrated in Fig.~\ref{fig:fig_1}. As shown in Fig.~\ref{fig:fig_1}(c), in the case of noncentrosymmetric Dirac points the energy bands are split when moving away from the fourfold band crossing, except however along the rotation axis, where a manifest twofold degeneracy is protected by point group symmetry. In contrast, in centrosymmetric systems with time-reversal symmetry all bands are manifestly twofold Kramers degenerate, giving rise to Dirac points shown in Fig.~\ref{fig:fig_1}(a). It is worth noting that, since the point groups $C_{4v}$ and $C_{6v}$ can be thought of as noncentrosymmetric descendants of $D_{4h}$ and $D_{6h}$, respectively, the breaking of inversion symmetry in centrosymmetric Dirac semimetals such as Cd$_{3}$As$_{2}$ and Na$_3$Bi does not necessarily split the Dirac points into twofold degenerate Weyl points shown in Fig.~\ref{fig:fig_1}(b). If either $C_{4v}$ or $C_{6v}$ symmetry are preserved, the fourfold Dirac points remain symmetry-protected, but are now of the kind shown in Fig.~\ref{fig:fig_1}(c). \par

\subsection{Model Hamiltonians \label{ssec:models}}

Based on these general symmetry arguments, our next step is to derive model Hamiltonians for each distinct type of band inversion. In systems with either $C_{4v}$ or $C_{6v}$ point group symmetry, the energy bands at $\Gamma$ can be labeled by their angular momentum quantum numbers $j_z$, which are in one-to-one correspondence with representations of the double point group. Due to the discrete nature of crystal rotation symmetry, the set of distinct angular momentum quantum numbers is finite and given by $j_z=\pm \frac12,\pm \frac32$ (fourfold rotation symmetry) and $j_z=\pm \frac12,\pm \frac32, \pm \frac52$ (sixfold rotation symmetry). To construct a minimal model for a band inversion-induced Dirac semimetal, we consider two bands, a conduction band and a valence band, which must have different angular momentum quantum numbers (i.e., are of different symmetry type). To describe the electronic states of the two bands, we introduce two sets of electron operators for the conduction ($c_\bk$) and valence band ($f_\bk$) as
\be
c_\bk = \begin{pmatrix} c_{\bk \up} \\  c_{\bk \down} \end{pmatrix}, \qquad f_\bk = \begin{pmatrix} f_{\bk \up} \\  f_{\bk \down} \end{pmatrix},    \label{eq:c-f}
\ee
where $\up,\down$ refers to the spin degree of freedom of each band. Note that the rotation eigenvalues of the spin states $\ket{\up}$ and $\ket{\down}$ depend on the symmetry type of each band and, in particular, will be different for the two bands. It is useful to collect the conduction and valence band degrees of freedom in the four-component spinor $\Psi_\bk$ defined as
\be
\Psi_\bk = \begin{pmatrix} c_{\bk } \\  f_{\bk } \end{pmatrix}.    \label{eq:psi}
\ee
Note that since we consider a vacuum defined by a filled valence band and an empty conduction band, $f_{\bk }$ creates holes in the valence band and can be viewed as a creation operator with respect to the vacuum. In this sense, the band inversion model bears a formal resemblance to the problem of superconductivity. \par 

With this resemblance in mind, we write the Hamiltonian $H$ for the two bands as $H= \sum_\bk\Psi^\dagger_\bk H_\bk  \Psi_\bk$,
where $H_\bk$ has the block structure
\be
 H_\bk=\begin{pmatrix} h^c_\bk  & \Delta_\bk \\ \Delta^\dagger_\bk & h^v_\bk \end{pmatrix}. \label{eq:Hk}
\ee
Here $h^{c,v}_\bk$ is the Hamiltonian describing the valence ($v$) and conduction  ($c$) bands, and $\Delta_\bk$ captures the coupling between the bands. Each of these are matrices in spin space, and the construction of a model Hamiltonian proceeds by determining the form of the matrices $h^{c,v}_\bk$ and $\Delta_\bk$ for a given type of the conduction and valence bands. 

Let $g$ be a point group symmetry and $U_g$ its matrix representation in the Hilbert space of the two bands. Then the matrix $U_g$ takes the form
\be
U_g = \begin{pmatrix} V_{g} & \\ & W_{g} \end{pmatrix}, \label{Ug}
\ee
where $V_{g}$ and $W_{g}$ are the matrix representations in the conduction and valence bands, respectively, i.e., the conduction and valence band states transform under $g$ as
\be
g\,:\, c^\dagger_\bk \rightarrow c^\dagger_{g\bk} V_g , \quad g\,:\, f^\dagger_\bk \rightarrow f^\dagger_{g\bk} W_g .
\ee
The matrix representations $V_{g}$ and $W_{g}$ depend on the symmetry quantum numbers of the $ \up,\down $-states within each band (i.e., the symmetry type of the band). Table \ref{tab:matrix} collects the matrix representations of the generators of $C_{4v}$ and $C_{6v}$ for the distinct angular momentum quantum numbers. Given $U_g$, invariance of the Hamiltonian under $g$ is expressed as
\be
U_gH_\bk U^\dagger_g  = H_{g\bk}. \label{H-g}
\ee
For the valence and conduction bands this leads to the requirement
\be
V_g h^c_\bk V^\dagger_g  = h^c_{g\bk}, \quad  W_g h^v_\bk W^\dagger_g  = h^v_{g\bk}, \label{hcv-g}
\ee
and for the coupling matrix $\Delta_\bk$ this implies
\be
V_g \Delta_\bk W^\dagger_g  = \Delta_{g\bk}.  \label{Delta-g}
\ee

\begin{table}[t]
\centering 
\begin{ruledtabular}
\begin{tabular}{cccc}
Band Type  &  $C_{4z}$  & $C_{6z}$  & $M_x$ \\
\hline  \\ [-2.0ex]
$j_z=\frac12$ & $e^{-i\pi\sigma_z/4}$  &  $e^{-i\pi\sigma_z/6}$  &   $\mp i\sigma_x$   \\  [2pt]
$j_z=\frac32$ & $e^{-i3\pi\sigma_z/4}$  & $e^{-i3\pi\sigma_z/6}$ &  $\pm i\sigma_x $  \\ [2pt] 
$ j_z=\frac52$      & $ -e^{-i\pi\sigma_z/4} $ & $e^{-i5\pi\sigma_z/6}$ &  $\mp i\sigma_x $   
\end{tabular}
\end{ruledtabular}
\caption{Tabulated list of the matrix representations of the point group generators for given angular momentum $j_z$. The generators of $C_{nv}$ are $C_{nz}$ and $M_x$, with $n=4,6$. }
\label{tab:matrix} 
\end{table}

To examine the constraints implied by these equations it is useful to expand $h^{c,v}_\bk$ and $\Delta_\bk$ in the Pauli matrices. For the conduction and valence bands we can expand $h^{c,v}_\bk$ as
\be
h^{c,v}_\bk  = \pm \varepsilon_\bk + \bb^{c,v}_\bk \cdot \makebf{\sigma}, \label{h-expand}
\ee
where the scalar function $ \varepsilon_\bk$ and the vector functions $\bb^{c,v}_\bk$ are real. Time-reversal symmetry $\Theta$ further requires that $\varepsilon_\bk$ is an even function of $\bk$ and that $\bb^{c,v}_\bk$ are odd. Here $+\varepsilon_\bk$ ($-\varepsilon_\bk$) corresponds to the conduction (valence) band, and for $\varepsilon_\bk$ we can take, up to quadratic order,
\be
\varepsilon_\bk = \varepsilon_0 + \frac{k^2_x+k^2_y}{2 m_x}+ \frac{k^2_z}{2 m_z},    \label{eq:epsilon}
\ee
where $m_x$ and $m_z$ are effective masses in the basal plane and $z$ direction, respectively. With this form of $h^{c,v}_\bk $, the energy separation of the bands is $2\varepsilon_0$ and $\varepsilon_0 < 0$ defines the band inverted regime. Note that in practice the conduction and valence bands can and will have different effective masses, but this difference is unimportant for our analysis. The lack of inversion symmetry allows for the second term in \eqref{h-expand}, parametrized by $\bb^{c,v}_\bk$; since $\bb^{c,v}_\bk$ is an odd function of $\bk$, inversion symmetry would force it to vanish. Physically it corresponds to Rashba-type spin-orbit splitting of the conduction and valence bands. Irrespective of band symmetry type, mirror symmetry $M_x \; :\; x\rightarrow -x$ imposes the general constraint
\be
b^x_\bk = b^x_{M_x\bk}, \quad -b^{y,z}_\bk = b^{y,z}_{M_x\bk}. \label{h-mirror}
\ee
Combined with invariance under rotations this constraint implies $b^z_\bk  =0$.

Next, consider the coupling matrix $\Delta_\bk$. We expand $\Delta_\bk$ as
\be
 \Delta_\bk  = \delta_\bk + \bd_\bk \cdot \makebf{\sigma}  \label{D-expand}
\ee
where the scalar function $\delta_\bk$ and the vector function $\bd_\bk$ are in general complex. Invariance under time-reversal $\Theta$ implies the condition
\be
 \delta^*_\bk = \delta_{-\bk} , \quad  -\bd^*_\bk = \bd_{-\bk}, \label{D-time}
\ee
for the (generally complex) functions $\delta_\bk$ and $\bd_\bk$. Furthermore, mirror symmetry $M_x$ imposes the contraints
\be
\pm \delta_\bk = \delta_{M_x\bk}, \quad \pm d^x_\bk = d^x_{M_x\bk}, \quad \mp d^{y,z}_\bk = d^{y,z}_{M_x\bk}, \label{D-mirror}
\ee
where the sign $\pm$ is determined by the absence or presence of a relative phase $e^{i\pi}$ in the matrix representations of the two bands (see Table \ref{tab:matrix}).

To determine the constraints from rotation symmetry imposed on $h^{c,v}_\bk $ and $\Delta_\bk$, we note that momentum $\bk$ transforms as
\beq
C_{nz} & : & k_\pm \rightarrow  e^{\pm i2\pi/n }k_\pm,   \label{k_Cnz} \\
M_{x} & : & k_\pm \rightarrow  -k_\mp ,  \label{k_Mx}
\eeq
where $k_\pm = k_x\pm ik_y$ and $n=4,6$ describes either a fourfold or sixfold crystal rotation. Momentum $k_z$ is left invariant by either symmetry. 

Based on these general considerations, we are now in a position to obtain the most general form of the Hamiltonian allowed by symmetry, for each type of band inversion. For each allowed coupling in the Hamiltonian, we only retain the lowest order contribution. \par

\subsection{Two examples of distinct band inversions \label{ssec:derivation}}

We apply the general formalism of the previous section to two particular cases, which will serve to illustrate all significant and relevant features of noncentrosymmetric Dirac semimetals. First, we consider a band inversion of $j_z=\pm\frac12$ and $j_z=\pm\frac32$ states, which we choose as the conduction and valence band, respectively. These states correspond to the representations $\Gamma_6$ and $\Gamma_7$ of $C_{4v}$, or $\Gamma_7$ and $\Gamma_9$ of $C_{6v}$. The form of $h^c_\bk$ does not depend on the degree of rotation symmetry, fourfold or sixfold, and is given by (see Appendix \ref{app:h_k})
\be
h^c_\bk = \varepsilon_\bk + \lambda_c(k_x\sigma_y - k_y\sigma_x). \label{eq:h^c}
\ee
The second term proportional to $\lambda_c$ is a Rashba spin-orbit coupling which linearly splits the Kramers pairs away from $\bk=0$. This ``intra-band'' spin-orbit coupling is associated with a characteristic momentum $m_z  \lambda_c$ or, alternatively, the energy scale $m_z  \lambda^2_c$, which may be compared to $\varepsilon_0$. 

The valence band Hamiltonian for the $j_z=\pm\frac32$ states does depend on the degree of rotation symmetry. In the case of fourfold rotational symmetry, which we will consider as an example here, $h^v_\bk$ is given by
\be
h^v_\bk = -\varepsilon_\bk + \lambda_v(k_x\sigma_y + k_y\sigma_x).  \label{eq:h^v}
\ee
The form of $h^v_\bk$ in the case of sixfold rotations is discussed in Appendix \ref{app:h_k}. 

We find that the coupling matrix $\Delta_\bk$ takes the form
\be
\Delta_\bk = \begin{pmatrix} \alpha  k_+ & \beta k^2_-+\beta' k^2_+ \\  -\beta k^2_+-\beta' k^2_- & \alpha  k_-\end{pmatrix},\label{eq:Delta-1/2-3/2}
\ee
where $\alpha, \beta, \beta'$ can be functions of $k_z$. In particular, we find that to lowest order in $k_z$, these coefficients are given by
\be
\alpha =  i  \alpha_{1} +  \alpha_{2}k_z, \quad \beta =  \beta_{1} + i \beta_{2}k_z,  \label{eq:a/b-1/2-3/2}
\ee
where $\alpha_{1,2}$ and $\beta_{1,2}$ are real; $\beta'$ has the same form as $\beta$. Here we have assumed a relative sign difference in the representation of mirror symmetry. The form of $\Delta_\bk$ in \eqref{eq:Delta-1/2-3/2} applies to both fourfold and sixfold rotational symmetry, except that $\beta'\equiv0$ for $C_{6v}$ symmetry. In the next section, we will examine the full Hamiltonian in detail and establish its phase diagram.

As a second example, we consider a band inversion of $j_z=\pm\frac12$ and $j_z=\pm\frac52$ states, which can only occur in materials with sixfold rotation symmetry. For this type of band inversion, the conduction and valence band Hamiltonians, $h^c_\bk$ and $h^v_\bk$, take the same form as Eqs.~\eqref{eq:h^c} and \eqref{eq:h^v}, respectively. Due to the different rotation eigenvalues of the valence band states, however, the form of the coupling matrix $\Delta_\bk$ is different, and is given by
\be
\Delta_\bk = \begin{pmatrix} \alpha  k^2_+ & \beta k^3_-+\beta' k^3_+ \\  -\beta k^3_+-\beta' k^3_- & \alpha  k^2_-\end{pmatrix}.  \label{eq:Delta-1/2-5/2}
\ee
The coefficients $\alpha, \beta, \beta'$ now take the form
\be
\alpha =   \alpha_{1} +  i \alpha_{2}k_z, \quad \beta = i \beta_{1} +  \beta_{2}k_z,  \label{eq:a/b-1/2-5/2}
\ee
where $\alpha_{1,2}$ and $\beta_{1,2}$ are real, as before, and $\beta'$ has the same form as $\beta$. Note that here we have assumed the same mirror representation for the two bands.

In the remainder of this section, and in Sec.~\ref{sec:analysis}, we will use the model Hamiltonians for these two types of band inversions to establish the generic properties of noncentrosymmetric Dirac semimetals.

\subsection{Mirror planes \label{ssec:mirror}}

As a first step in our analysis, we consider the mirror planes, where the Hamiltonian commutes with the mirror symmetry operator. As an example, consider the $k_x=0$ mirror plane, where $\bk = (k_y,k_z)$. Since the Hamiltonian commutes with $M_x$ it can be block-diagonalized by expressing it in terms of eigenstates of $M_x$, resulting in two diagonal blocks labeled by the mirror eigenvalues $\pm i$. The Hamiltonian $H^{\pm i}_\bk$ for each block can be expanded as
\be
H^{\pm i}_\bk = \chi^\pm_\bk + \bn^\pm_\bk \cdot \makebf{\tau}, \label{eq:H-mirror}
\ee
where $\makebf{\tau} =(\tau_x,\tau_y ,\tau_z)$ is a set of Pauli matrices. The product of time-reversal symmetry and the twofold rotation $\Theta C_{2z}$ anti-commutes with $M_x$, i.e., $\{M_x,\Theta C_{2z}  \} =0$, and imposes the constraint 
\be
(H^{\pm i})^*_{k_yk_z} = H^{\pm i}_{k_y,-k_z}, \label{eq:mirror-ky}
\ee
which implies that $n^{y,\pm}$ is an odd function of $k_z$. In particular, it implies that on the $k_y$ axis, where $k_z=0$ (and, of course, $k_x=0$), we must have $n^{y,\pm}=0$. This is an important property which will be exploited in Secs.~\ref{ssec:weyl_point} and \ref{ssec:weyl_line} when we address gap closing transitions.

It is straightforward to obtain $H^{\pm i}_\bk $ for the two types of band inversions considered in Sec.~\ref{ssec:derivation}. For the first type, between $j_z=\pm\frac12$ and $j_z=\pm\frac32$ states, we find 
\be
 \chi^\pm_\bk  = \pm \lambda_1 k_y , \quad n^{z,\pm}_\bk = \varepsilon_\bk \pm \lambda_2 k_y, \label{mirror_C4_1}
\ee
where $\lambda_{1,2} = (\lambda_c\pm \lambda_v)/2$. For the components $n^{x,y}_\bk$ we find
\beq
n^{x,\pm}_\bk &=& -\alpha_1 k_y \mp (\beta_1 + \beta'_1)k^2_y , \label{mirror_C4_2} \\
n^{y,\pm}_\bk &=& -\alpha_2 k_z k_y \pm (\beta_2 + \beta'_2) k_z k^2_y  \label{mirror_C4_3}.
\eeq
Observe that $n^{y,\pm}_\bk$ is indeed an odd function of $k_z$, as required. Within each mirror sector the two branches of the energy spectrum are easily obtained as 
\be
\mathcal E^\pm_{\bk s} = \chi^\pm_\bk  + s |\bn^\pm_\bk|, \label{eq:Ek-mirror}
\ee
where $s=\pm 1$. For $k_y\rightarrow 0$ the $+$ ($-$) solution can be identified with the conduction (valence) band, whereas for large $k_y$ this is reversed.


For the second type of band inversion, between $j_z=\pm\frac12$ and $j_z=\pm\frac52$ bands, and with $\Delta_\bk$ given by Eq.~\eqref{eq:Delta-1/2-5/2}, we find 
\be
 \chi^\pm_\bk  = \pm \lambda_2 k_y , \quad n^{z,\pm}_\bk = \varepsilon_\bk \pm \lambda_1 k_y,  \label{mirror_C6_1}
\ee
as well as
\beq
n^{x,\pm}_\bk &=& -\alpha_1 k^2_y \pm (\beta_1 - \beta'_1)k^3_y , \label{mirror_C6_2}\\
n^{y,\pm}_\bk &=& \alpha_2 k_zk^2_y \pm (\beta_2 - \beta'_2)k_zk^3_y  \label{mirror_C6_3}.
\eeq
The energy spectrum is obtained as before and given by Eq.~\eqref{eq:Ek-mirror}.

We conclude this section by stressing that here we have only considered one particular mirror plane ($k_x=0$) as an example. This mirror plane belongs to set of two (in the case of $C_{4v}$) or three (in the case of $C_{6v}$) equivalent mirror planes, which correspond to conjugate mirror symmetries. As a result, the energy spectrum must be identical on all equivalent mirror planes. Importantly, there is second set of mirror planes, {\it not} related to the first by similarity transformation, for which a similar analysis can be performed. In the case of $C_{6v}$, the $k_y=0$ plane belongs to the second set and the Hamiltonian is block diagonal in a basis of $M_y$ eigenstates. Since the Hamiltonian is not conjugate to the Hamiltonian on the $k_x=0$ plane, its spectrum is generally not identical. This will be of consequence when, in the next section, we consider the creation and annihilation of Weyl points, which are created on one set of mirror planes and annihilated on the other. Furthermore, we will show that Weyl lines nodes can occur on either one of the two sets of mirror planes, but not simultaneously. Weyl lines nodes---if present---can only be realized on one of the two sets.

\section{Phase diagram of noncentrosymmetric Dirac semimetals \label{sec:analysis}}

Based on the obtained model Hamiltonians, we now examine the general phase diagram of noncentrosymmetric Dirac semimetals. In particular, our goal is to establish the distinct topological semimetallic phases which can generically occur for a particular type of band inversion. To achieve this, we first focus on the region close to the Dirac points and examine the dispersion in the vicinity of the Dirac band crossings. This allows us to determine how the qualitative structure of the Dirac points changes as a function of model parameters, and how this gives rise to different types of Dirac point crossings in noncentrosymmetric systems. We then focus on the vertical mirror planes and study how distinct semimetallic phases arise in the phase diagram, separated by gap closing transitions.

\subsection{Description of the Dirac points \label{ssec:dirac}}

By construction, the model Hamiltonians derived in the previous section describe a set of two fourfold degenerate band crossings on the rotation axis. These crossings occur at $k_z=\pm K_0$, with $K_0 = \sqrt{2m_z|\varepsilon_0|}$, and realize Dirac points. Since our goal is to determine the precise nature of the noncentrosymmetric Dirac points, we expand the dispersion at the Dirac points in momentum $\bq = \bk \mp \bK_0$, where $\bK_0 = (0,0, K_0)$. For concreteness, we will focus on the Dirac point at $+\bK_0 $; the results for $-\bK_0 $ are straightforward and similar. 

Stable Dirac points are defined as a symmetry-protected degeneracy of two Weyl points of opposite chirality (i.e., handedness). To expose this structure in the present context, it is useful to consider the inversion symmetric limit of the model Hamiltonian. Two possible limits exist, one where the two bands have equal parity and one where the bands have opposite parity. As we show in more detail in Appendix~\ref{app:inversion}, starting from either case it is possible to bring the Hamiltonian describing the electronic states near $\bK_0$, denoted $\mathcal H_\bq$, into the form
\be
\mathcal H_\bq= \begin{pmatrix} A^-_\bq & B_\bq \\  B^\dagger_\bq & A^+_\bq \end{pmatrix} . \label{eq:H_q}
\ee
Here $A^\pm_\bq$ describes a (twofold degenerate) Weyl node with positive ($+$) and negative ($-$) chirality, and $B_\bq$ describes a coupling between the two Weyl nodes, which is allowed due to the lack of inversion symmetry. We now examine the form of $A^\pm_\bq$ and $B_\bq$ for the two representative types of band inversions discussed above in Sec.~\ref{ssec:derivation}. 

Consider first the case of a band inversion of $j_z=\pm \frac12$ and $j_z=\pm \frac32$ states, as defined via Eqs.~\eqref{eq:h^c}--\eqref{eq:Delta-1/2-3/2}. In this case the two Weyl nodes are described by
\be
A^\pm_\bq= q_iA^\pm_{ij}\tau_j, \label{eq:A_q_pm}
\ee
with matrices $A^\pm$ given by
\be
A^\pm = \begin{pmatrix} \tilde \alpha_2 & -\alpha_1 & 0 \\ \pm \alpha_1  &\pm \tilde \alpha_2& 0\\ 0 &0& v_z \end{pmatrix}  \label{eq:A_pm_1/2-3/2},
\ee
and velocities $v_z$ and $\tilde \alpha_2$ defined as $v_z \equiv K_0/m_z$ and $\tilde \alpha_2  \equiv \alpha_2 K_0$.
As required, Eq.~\eqref{eq:A_q_pm} has indeed the general form of a linear crossing of energy bands, and the matrices $A^\pm$ encode the Weyl node chirality through the sign of their determinant, i.e., $\text{Det}\,A^\pm = \pm v_z(\alpha^2_1 + \tilde \alpha^2_2 )$. Two results from follow from this: first, from Eq.~\eqref{eq:A_pm_1/2-3/2} we see that, apart from the velocity $v_z$, the matrices $A^\pm$ only depend on the couplings between the bands captured by the off-diagonal block $\Delta_\bk$; they do not depend on the intra-band spin-orbit couplings described by $h^{c,v}_\bk$. Second, the nature of the Weyl points does not change as a function of parameters; it never vanishes and does not change sign.

\begin{figure}[pt]
    \centering
    \includegraphics[width=\columnwidth]{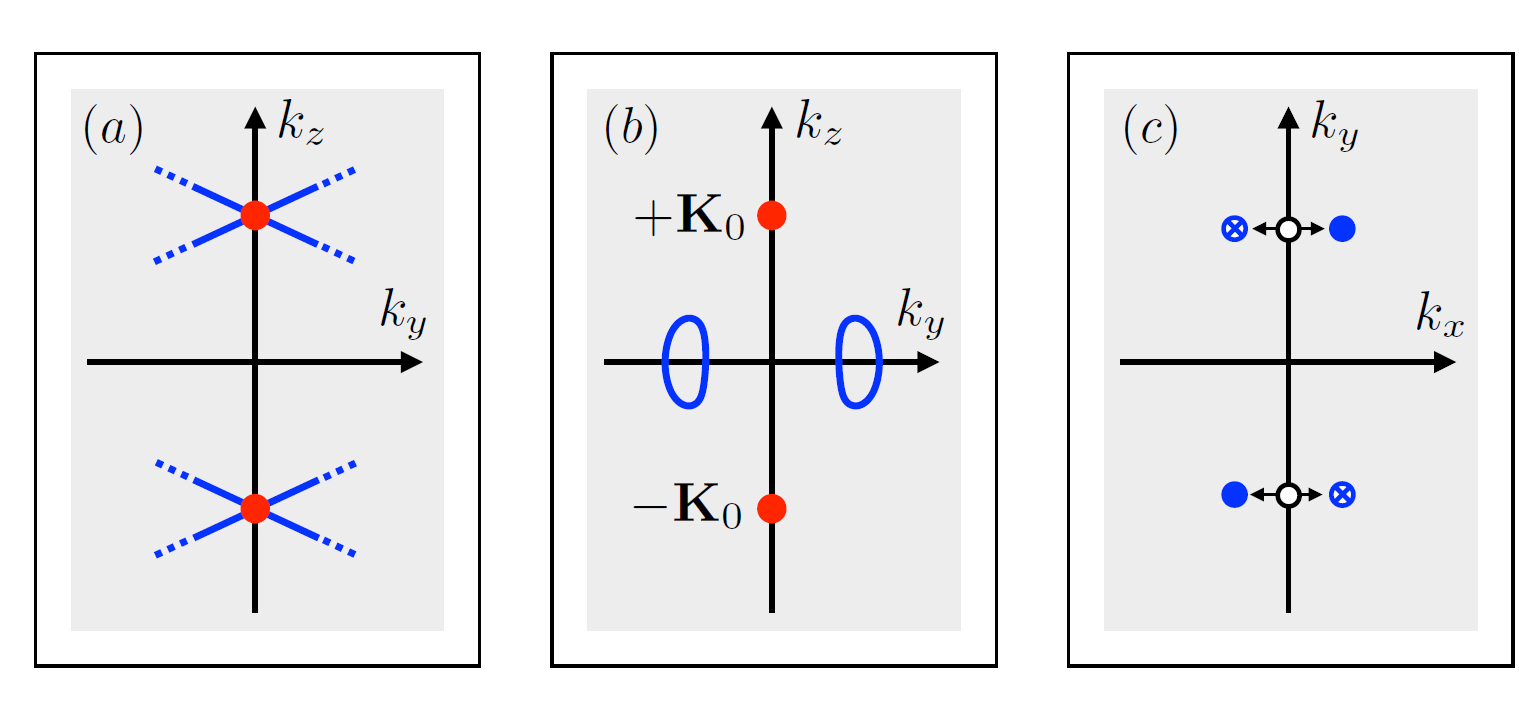}
    \caption{Schematic representation of the creation of Weyl line nodes and Weyl point nodes. Panel (a) and (b) show the $k_y$-$k_z$ plane, with the Dirac points at $\pm \bK_0$ on the $k_z$ indicated by bold red dots. As discussed in Sec.~\ref{ssec:dirac} and shown in (a), inversion symmetry breaking can give rise to Weyl line nodes which connect to the Dirac points, realizing a qualitatively different type of Dirac point. Panel (b) and (c) show gap closing transitions on the $k_y$ axis, which can give rise to Weyl line nodes (panel b) or Weyl point nodes (panel c), depending on the mirror eigenvalues of the bands (see Secs.~\ref{ssec:weyl_point} and \ref{ssec:weyl_line}). }
    \label{fig:weyl}
\end{figure}

In contrast, $B_\bq$ is given by
\be
B_\bq = -i\lambda_1 q_x-\lambda_2q_y -(i\lambda_2 q_x+\lambda_1q_y)\tau_z,  \label{eq:B_q}
\ee
and only depends on the Rashba-type intra-band spin-orbit couplings. Recall that $\lambda_{1,2} = (\lambda_c \pm \lambda_v)/2$. The structure of the Dirac point may change as function of the coupling described by $B_\bq$ and we thus need to examine the effect of $B_\bq$. The possible effect of $B_\bq$ is to cause energy level crossings away from the Dirac point, and as a result, we can restrict the analysis to the mirror planes. Energy level crossings can only occur on the mirror planes, provided the energy levels have different mirror eigenvalues. Such band crossings, if they occur, occur on a line in momentum space and give rise to topological Weyl line nodes. This is shown schematically in Fig.~\ref{fig:weyl}(a), where the bold blue lines indicate lines of degeneracy. Since these line nodes connect to the Dirac points, the nature of the Dirac point qualitatively changes \cite{Gao16p205109}. 

Taking the $yz$ mirror plane as an example, we set $q_x=0$ and transform to the mirror eigenbasis. By definition, this yields a Hamiltonian exactly equal to the Hamiltonian obtained by expanding Eqs.~\eqref{mirror_C4_1}--\eqref{mirror_C4_3} to linear order in $(q_y,q_z)$. By setting the energies corresponding to different mirror eigenvalues equal, we obtain a  condition for the existence of line nodes, which is given by
\be
 \lambda^2_1-\lambda^2_2-\alpha^2_1 - \tilde\alpha^2_2 >0. \label{eq:linenode_condition}
\ee
This condition leads to the important conclusion that line nodes exist for sufficiently large $\lambda_1 = (\lambda_c+\lambda_2)/2$, which is directly related to the strength of the spin-orbit splitting of the conduction and valence bands. As a result, for this type of band inversion two kinds of Dirac points can occur: {\it (i)} conventional Dirac points with linearly dispersing (non-degenerate) bands away from the fourfold crossing at $\pm\bK_0$, and {\it (ii)} unconventional Dirac points at which symmetry-protected line nodes terminate, as shown in Fig.~\ref{fig:weyl}(a). Our analysis shows that this distinction is determined by the relative strength of intra-band spin-orbit coupling and the coupling between the bands. 

Two remarks are in order. First, since we start from an expansion close to the Dirac point, see Eq.~\eqref{eq:H_q}, it is important to note that the analysis presented here applies only close to the Dirac points. The fate of the line nodes away from the Dirac point cannot be determined [as is indicated by dashed blue lines in Fig.~\ref{fig:weyl}(a)], but will be addressed in Sec.~\ref{ssec:weyl_line}. Second, we note that the line nodes occur on all symmetry-equivalent mirror planes, i.e., mirror planes related by rotation. As mentioned in Sec.~\ref{ssec:mirror}, there are two sets of such symmetry-equivalent mirror planes and by performing the analysis leading to \eqref{eq:linenode_condition} for both sets, we find that line nodes can occur on either set of mirror planes, depending on parameters, but not simultaneously; the existence of Weyl line nodes on the two sets of inequivalent mirror planes is mutually exclusive.

Next, consider the second type of band inversion discussed in the previous section, involving $j_z=\pm \frac12$ and $j_z=\pm \frac52$ states and thus requiring sixfold rotation symmetry. We find that the nature of the Dirac points in this case is rather different. The key difference is manifest in the form of $A^\pm_\bq$, which are given by
\be
A^\pm_\bq = v_z q_z\tau_z  +  \bar \alpha q^2_\mp\tau_+ +   \bar \alpha^* q^2_\pm \tau_-.  \label{eq:A_pm_1/2-5/2}
\ee
Here, we have again focused on the Dirac point at $+\bK_0$, and further defined $\tau_\pm  = (\tau_x \pm i \tau_y)/2$, as well as $\bar \alpha = \alpha_1+i\alpha_2K_0$. Importantly, in this case $A^\pm_\bq$ describe a Weyl node with quadratic dispersion in $(q_x,q_y)$ and Berry monopole charge $C=\pm2$~\cite{Fang12p266802}. The form of $B_\bq$ is the same as in Eq.~\eqref{eq:B_q}, which in particular implies that $B_\bq$ introduces terms linearly dependent on $(q_x,q_y)$. Since the latter are more important for small $\bq$, we may ignore all quadratic terms and retain only the linear contributions to $\mathcal H_\bq$. It is then straightforward to demonstrate that line nodes terminating at the Dirac points \emph{always} exist on one of the two sets of inequivalent mirror planes. In particular, the condition for line nodes on the $yz$ and symmetry-related mirror planes is $|\lambda_2| > |\lambda_1| $, whereas on the $xz$ and symmetry-related planes it is $|\lambda_1| > |\lambda_2| $; one of these conditions is always satisfied.

We summarize our analysis in the vicinity of the Dirac points by concluding that the way in which the nature of topological Dirac points is affected by inversion symmetry breaking crucially depends on the type of band inversion. In particular, Dirac points which may be viewed as composed of two $C=\pm 2$ Weyl points necessarily appear in combination with Weyl line nodes. Weyl line nodes may also occur for Dirac points with purely linear dispersion, but in that case the presence of line nodes is a threshold phenomenon: it depends on the strength of the intra-band spin-orbit splitting.

\begin{figure}[pt]
    \centering
    \includegraphics[width=\columnwidth]{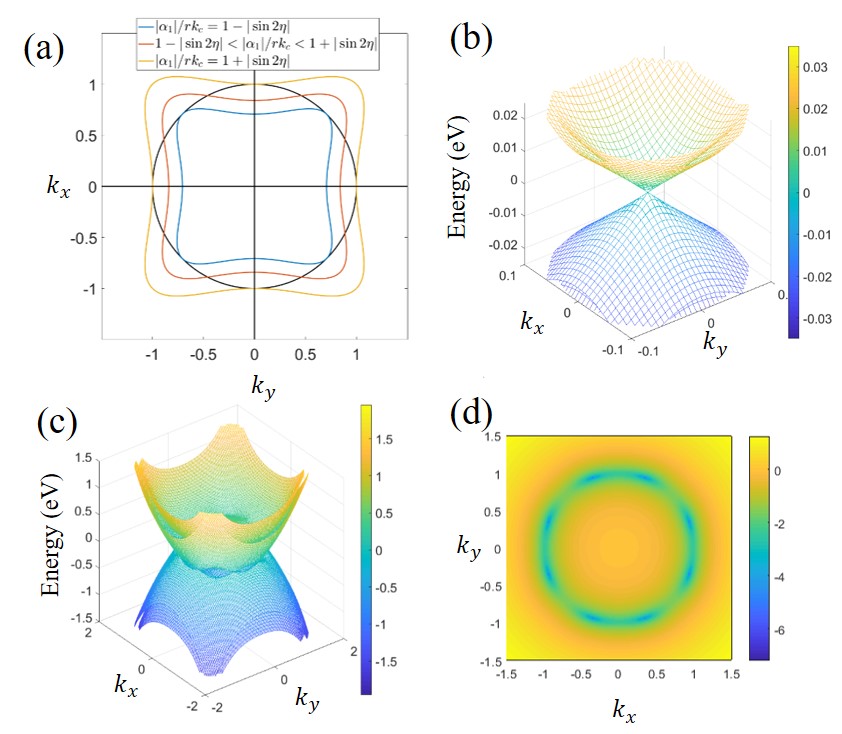}
   \caption{Presence of Weyl points in a crystal with $C_{4v}$ point group symmetry and a band inversion of $j_z=\pm \frac12$ and $j_z=\pm \frac32$ states. (a) The phase diagram of creation and annihilation of Weyl points in the $k_z=0$ as described by the plots of $1+\sin 2\eta \cos4\theta_\bk = |\alpha_1|/rk_c$ and circle  $k=k_c$. (b) Three dimensional band structures of Dirac points with the band dispersions along $k_{x}$ and $k_{y}$ directions. (c) Three dimensional band structures on the $k_{z}=0$ plane. (d) The colormap of energy difference with logarithmic scale between the conduction and valence bands on the $k_{z}=0$ plane.}
    \label{fig:fig_2}
\end{figure}

\subsection{Gap closings on the mirror planes: Weyl point nodes \label{ssec:weyl_point}}

The next step in our analysis of the generic phase diagram is to study the occurrence of gap closings on the mirror planes. Since on the mirror plane all states can be labeled by the mirror eigenvalues, two types of gap closings are possible: between two (non-degenerate) bands with ({\it i})  equal mirror eigenvalues or ({\it ii}) opposite mirror eigenvalues.~\cite{Murakami07p356,Yang13p086402,Liu14p155316,Murakami17p1602680} In case ({\it i}) , a gap closing is known to give rise to the nucleation of two Weyl points, as schematically illustrated in Fig.~\ref{fig:weyl}(c), whereas case ({\it ii}) leads to the creation of stable Weyl line nodes, shown in Fig.~\ref{fig:weyl}(b). Here we systematically investigate both possibilities within the context of the model Hamiltonians.

We first consider case ({\it i}) and defer case ({\it ii}) to Sec.~\ref{ssec:weyl_line}. Two planes in momentum space are important for the creation of Weyl points: the mirror plane and the $k_z=0$ plane. On the $k_z=0$ plane the combined $\Theta C_{2z}$ symmetry ensures the local stability of Weyl points~\cite{Soluyanov15p495,Fang15p161105} and as a result, nucleation of Weyl points can only occur on the intersection of these two planes, as is clear from Eq.~\eqref{eq:mirror-ky}.\cite{Murakami17p1602680} Let us focus on one of such intersections, the $k_y$ axis. Since $n^{y,\pm}=0$ on the $k_y$ axis, a gap closing will generically occur for some $k_{y0}$ by changing one parameter of the Hamiltonian. More formally, we may view the mirror-diagonal Hamiltonian $H^{\pm i}(k_y,m)$ as a function of some gap closing parameter $m$, such that at $m=m_0$ the gap closes at $k_{y0}$. The gap closing condition is then expressed as
\be
n^{x}(k_{y},m) = n^{z}(k_{y},m)=0,
\ee
with solutions $(k_{y0},m_0)$.

In the case of a band inversion of $j_z=\pm \frac12$ and $j_z=\pm \frac32$ states described by $\Delta_\bk$ in Eq.~\eqref{eq:Delta-1/2-3/2}, the gap closing condition takes the specific form
\beq
0 & = & \varepsilon_0 + \frac{k^2_y}{2m_x} \pm \lambda_2 k_y, \label{gap-close_1} \\
0 & = & -\alpha_1 k_y \mp (\beta_1 + \beta'_1)k^2_y. \label{gap-close_2}
\eeq
Note first that this set of equations can only have solutions due to the breaking of inversion symmetry, since lack of inversion symmetry allows both $\alpha_1$ and $\beta^{(\prime)}_1$ to be nonzero. Since we have $\varepsilon_0 <0$, Eq.~\eqref{gap-close_1} always has two solutions for $k_y$, and by designating one of the parameters of Eq.~\eqref{gap-close_2} the gap closing parameter and changing it continuously, \eqref{gap-close_2} can be satisfied at one of the zeros of \eqref{gap-close_1}. Alternatively, the Rashba spin-orbit coupling strength $\lambda_2$ may be viewed as the gap closing parameter. From Eqs.~\eqref{gap-close_1} and \eqref{gap-close_2} we can further conclude that if a gap closing occurs at $k_{y0}$ in the $+ i$ mirror sector, then a simultaneous gap closing occurs at $-k_{y0}$ in the $- i$ mirror sector. This follows from the invariance of the equations under simultaneously changing $k_y \rightarrow -k_y$ and $+ \rightarrow -$. As a result, when the starting point is a fully gapped $k_z=0$ plane, two pairs of Weyl points are created on the mirror plane at the gap closing transition. As the gap closing parameter is further changed, the Weyl points belonging to each pair move in opposite directions perpendicular to the mirror plane, as shown schematically in Fig.~\ref{fig:weyl}(c). A similar gap closing analysis may be performed for a band inversion of $j_z=\pm \frac12$ and $j_z=\pm \frac52$ states based on the mirror plane Hamiltonian \eqref{mirror_C6_1}--\eqref{mirror_C6_3}.

The resulting phase, which not only exhibits Dirac points on the rotation axis but also Weyl points on the $k_z=0$ plane, defines a new type of topological semimetal with coexistent Dirac and Weyl fermions, which was first proposed in Ref.~\onlinecite{Gao18p106404}. As a function of Hamiltonian parameters, pairs of Weyl points can be created on one set of mirror planes related by rotation symmetry (see Sec.~\ref{ssec:mirror}), and may exchange partners by annihilating pairwise on the other (and inequivalent) set of mirror planes. This defines a topological transition within the two-dimensional $k_z=0$ plane~\cite{Murakami07p356,Murakami08p165313,Gao18p106404}, leading to a change of the corresponding $\mathbb{Z}_2$ index.\cite{Kane05p146802}

\begin{figure}[pt]
    \centering
    \includegraphics[width=\columnwidth]{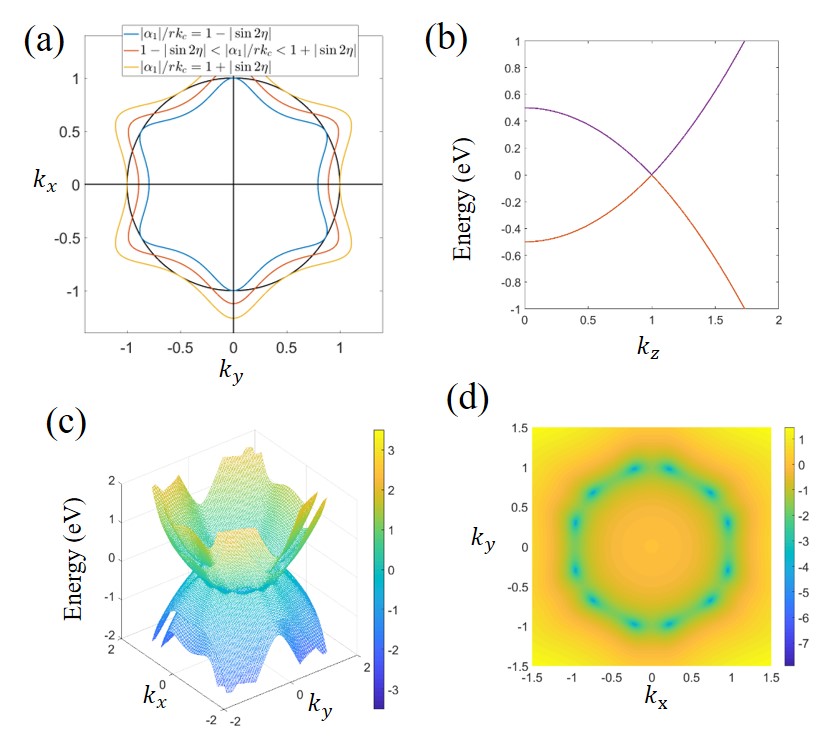}
    \caption{Phase diagram of Weyl points and electronic structures of crystals in $C_{6v}$ with $\Gamma_{7}$ and $\Gamma_{8}$ bands from $k \cdot p$ model. 
    (a) The phase diagram of creation and annihilation of Weyl points in the $k_z=0$ as described by the plots of $1+\sin 2\eta \cos6\theta_\bk = |\alpha_1|/rk_c$ and the circle $k=k_c$. (b) The band dispersions along $k_{z}$ directions. (c) 3D band structures on the $k_{z}=0$. (d) The colormap of energy difference with logarithmic scale between the conduction and valence band on the $k_{z}=0$ plane.}
    \label{fig:fig_3}
\end{figure}

To study the creation and annihilation of Weyl points in the $k_z=0$ plane in more detail, we go beyond the gap closing analysis on the mirror invariant line and solve for the full spectrum on the $k_z=0$ plane. Since the existence of Weyl points does not rely on the Rashba spin-orbit band splitting, we make the simplifying assumption $\lambda_v=\lambda_c=0$. As shown in Appendix \ref{app:diag}, the Hamiltonian is straightforwardly diagonalized and in the case of $\Delta_\bk$ given by Eq.~\eqref{eq:Delta-1/2-3/2}, we find the energies
\be
E^2_\bk = \varepsilon_\bk^{2} + \left(\sqrt{k^4r^2(1+\sin 2\eta \cos4\theta_\bk)} \pm \alpha_1 k\right)^2, \label{eq:Ek_C4}
\ee
where we have defined $r^2\equiv \beta^2_1+{\beta'}^2_1$ and $\tan \eta \equiv \beta'_1/\beta_1$ as a parametrization of the coupling constants $\beta_1,\beta'_1$. We have further defined $k^2 \equiv k^2_x+k^2_y$ and $\theta_\bk \equiv \arctan (k_y/k_x)$. (Note that $\bk$ is restricted to the $k_z=0$ plane.) The spectrum defined by Eq.~\eqref{eq:Ek_C4} features Weyl points whenever $E_\bk=0$ has solutions, which can only occur for two of the four spectral branches and must occur on the circle defined by $k_c = \sqrt{2|\varepsilon_0|m_x}$. The angle $\theta^c_\bk$ at which the crossing occurs is given by the solution of the equation $1+\sin 2\eta \cos4\theta_\bk = |\alpha_1|/rk_c$, from which we conclude that Weyl points exist when the two conditions $|\alpha_1|/rk_c < 1+|\sin 2\eta |$ and $|\alpha_1|/rk_c > 1-|\sin 2\eta |$ are satisfied. Note that the sign of $\sin 2\eta$ depends on the relative sign of $\beta_1$ and $\beta'_1$: for $\beta_1\beta'_1>0$ ($\beta_1\beta'_1<0$) one has $\sin 2\eta > 0$ ($\sin 2\eta < 0$). This determines on which set of mirror planes the Weyl points are created and annihilated as parameters are changed.

The creation and annihilation of Weyl points on the $k_z=0$ plane of a fourfold rotation symmetric system is shown in Fig.~\ref{fig:fig_2}, based on the energy solutions of \eqref{eq:Ek_C4}. In particular, Fig.~\ref{fig:fig_2}(a) illustrates the range of values of $|\alpha_1|/r$ for which the Weyl points are realized in the $k_z=0$ plane. The Weyl points are located where the orange curve crosses the black circle with radius $k/k_c=1$. The yellow and blue curves correspond to the critical values $|\alpha_1|/r = k_c(1+ |\sin2\eta|)$ and $|\alpha_1|/r = k_c(1- |\sin2\eta|)$, respectively, at which creation and annihilation occurs. Figure \ref{fig:fig_2}(d) shows the energy difference between the conduction and valence bands on a logarithmic scale, clearly revealing the existence of pairs of Weyl points. Furthermore, Fig.~\ref{fig:fig_2}(d) shows that the linear dispersion away from the Weyl points is very anisotropic. We indeed expect a much smaller velocity in the direction tangential to the circle $k/k_c=1$ as compared to the perpendicular direction. 

A similar analysis of Weyl points can be performed for a band inversion described by Eq.~\eqref{eq:Delta-1/2-5/2}, in which case we find the energies
\be
E^2_\bk = \varepsilon_\bk^{2} + \left(\sqrt{k^6r^2(1+\sin 2\eta \cos6\theta_\bk)} \pm \alpha_1 k^2\right)^2. \label{eq:Ek_C6}
\ee
These solutions have a similar structure but exhibit a sixfold crystalline anisotropy, instead of fourfold, due to the sixfold rotation symmetry of $H_\bk$ defined by Eq.~\eqref{eq:Delta-1/2-5/2}. In particular, Weyl points occur whenever the equation $1+\sin 2\eta \cos6\theta_\bk = |\alpha_1|/rk_c$ has solutions, with $k_c = \sqrt{2|\varepsilon_0|m_x}$ as before. 

An illustration of the emergence of Weyl points in a sixfold symmetric system is shown in Fig.~\ref{fig:fig_3}. Sixfold rotation symmetry gives rise to six pairs of Weyl points, as shown in Fig.~\ref{fig:fig_3}(d).

\subsection{Gap closings on the mirror planes: Weyl line nodes\label{ssec:weyl_line}}

We now turn to the emergence of Weyl line nodes on the mirror planes. As mentioned, Weyl line nodes are realized when bands with opposite mirror eigenvalue cross, thus preventing a coupling between the bands. In Sec.~\ref{ssec:mirror}, we block diagonalized the Hamiltonian on the mirror planes using the mirror eigenstates and this form of the Hamiltonian, in particular the mirror-resolved energy spectrum of Eq.~\eqref{eq:Ek-mirror}, form the basis of our analysis. Since lines nodes are defined as crossings of bands with opposite mirror eigenvalue, they are solutions to the equation
\be
0 = \mathcal E^+_{\bk, +} - \mathcal E^-_{\bk, -}  = \mathcal E^-_{\bk ,+} - \mathcal E^+_{\bk, -}, \label{eq:weyl_line_condition}
\ee
where $\bk$ is understood to be restricted to the mirror plane. 

In the present class of models, Weyl line nodes can be created in two different ways as function of parameters. First, as discussed in detail in Sec.~\ref{ssec:dirac}, line nodes can emerge in the vicinity of the Dirac points, see also Fig.~\ref{fig:weyl}(a). Second, line nodes can emerge after a gap closing on the intersection of the (vertical) mirror plane and the $k_z=0$ plane~\cite{Murakami17p1602680}, which is schematically illustrated in Fig.~\ref{fig:weyl}(b). As our analysis in Sec.~\ref{ssec:dirac} has shown, the Weyl line nodes which connect to the Dirac points can only occur for nonzero and sufficiently large intra-band spin orbit coupling $\lambda_1$ and $\lambda_2$. This is true for Weyl line nodes in general: the existence of line nodes on the mirror plane requires nonzero $\lambda_1,\lambda_2$. To see this in a more general setting, consider the block diagonal Hamiltonians obtained in Sec.~\ref{ssec:mirror}, which correspond to the $k_y$--$k_z$ mirror planes. Setting $k_z=0$ and $\lambda_1=\lambda_2=0$, it is straightforward to show that Eq.~\eqref{eq:weyl_line_condition} does not have solutions. Hence, our first key result is that Weyl line nodes originate from strong intra-band Rashba-type spin-orbit coupling.

As stressed in Sec.~\ref{ssec:mirror}, systems with point group symmetry $C_{4v}$ or $C_{6v}$ have two inequivalent sets of mirror planes. Due to this inequivalence, the question arises whether the presence of line nodes on one set is correlated with line nodes on the other set. We find that this is indeed the case: Weyl line nodes can only occur on one of the two inequivalent sets of mirror planes. On which set they occur---if they occur---depends on whether the ratio $|\lambda_1|/|\lambda_2|$ is smaller or larger than one. This is due to the fact that the role of $\lambda_1$ and $\lambda_2$ is reversed on the two sets of mirror planes, which is straightforwardly established. 

\begin{figure}[pt]
    \centering
    \includegraphics[width=\columnwidth]{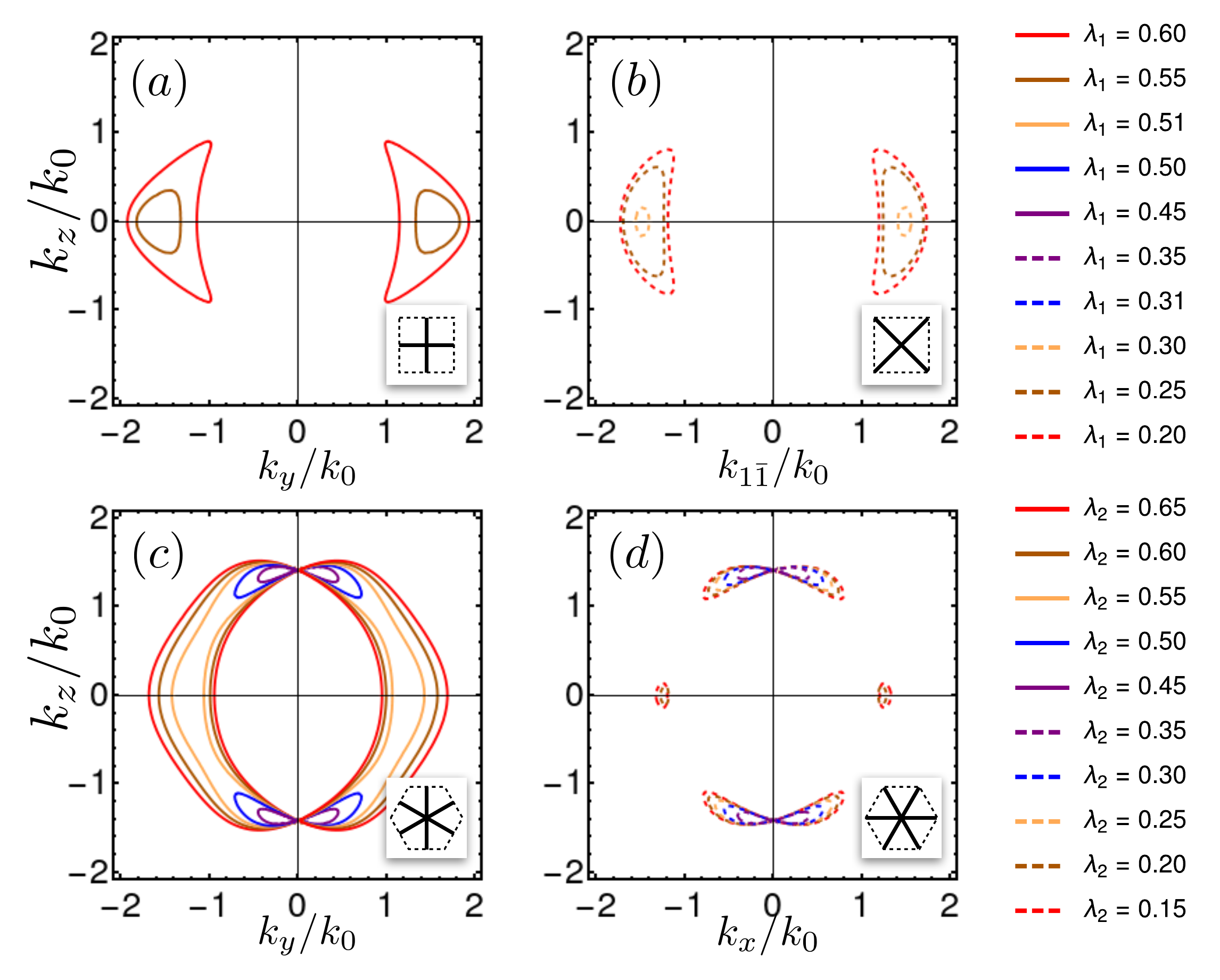}
    \caption{Emergence of Weyl nodes on the mirror planes. Panels (a) and (b) show lines nodes of a fourfold rotation symmetric model for a band inversion between $j_z=\pm \tfrac12$ and $j_z=\pm \tfrac32$ bands, where as panels (c) and (d) show the lines nodes of model for a band inversion between $j_z=\pm \tfrac12$ and $j_z=\pm \tfrac52$ bands. The two columns, i.e., panels (a)-(c) and (b)-(d), show results for the two inequivalent sets of mirror planes, which are schematically indicated by bold lines in the bottom corner insets. Different curves correspond to different values of $\lambda_1$ or $\lambda_2$, both measured in units $v = k_0/m$, with $m \equiv m_x=m_z$ and unit of momentum $k_0=\sqrt{|\varepsilon_0|m}$. Results of (a)-(b) are obtained by solving Eq.~\eqref{eq:weyl_line_condition} using the Hamiltonians defined via \eqref{eq:a/b-1/2-3/2} with parameters $\alpha_1/v= 0.26$, $m \alpha_2 = 0.28$, $m \beta_1 = 0.12$, $m \beta'_1 = 0.08$, $m k_0 \beta_2 = 0.06$, $m k_0 \beta'_2 = 0.10$; and results of (c)-(d) are obtained using \eqref{eq:a/b-1/2-5/2} (c-d) with the same parameters (albeit different units). We have set $\lambda_2/v=0.4$ and $\lambda_1/v=0.4$ in (a)-(b) and (c)-(d), respectively.}
    \label{fig:weyl_line}
\end{figure}

We demonstrate these results by showing the solutions of Eq.~\eqref{eq:weyl_line_condition} for different values of $\lambda_1,\lambda_2$ in Fig.~\ref{fig:weyl_line}.
Consider first panels (a) and (b) of Fig.~\ref{fig:weyl_line}, which are obtained for an inversion of $j_z=\pm \tfrac12$ and $j_z=\pm \tfrac32$ bands in a $C_{4v}$ symmetric system. Panel (a) shows the solutions on the $k_x=0$ mirror plane, whereas panel (b) shows the solutions $k_x+k_y=0$ mirror plane, for different values of $\lambda_1/v$, keeping $\lambda_2/v=0.4$ fixed. Here $v=k_0/m_x$ with the unit of momentum defined as $k_0=\sqrt{|\varepsilon_0|m_x}$. As demonstrated by Fig.~\ref{fig:weyl_line}(a) and (b), Weyl line nodes only occur on the $k_x=0$ ($k_x+k_y=0$) mirror plane for $\lambda_1 > \lambda_2$ ($\lambda_1 < \lambda_2$). Note further that for the chosen set of parameters (see Figure caption) there are no line nodes connecting to the Dirac points. This is markedly different in panels (c) and (d), which show the solutions of Eq.~\eqref{eq:weyl_line_condition} for a band inversion of $j_z=\pm \tfrac12$ and $j_z=\pm \tfrac52$ bands. (Recall that this requires sixfold rotation symmetry.) As is evident, line nodes connecting to the Dirac points are present for all values of $\lambda_2/v$ on one of the two sets of mirror planes, consistent with our analysis of Sec.~\ref{ssec:dirac}. Note, however, that the line nodes exist only on one set of mirror planes, depending on whether $\lambda_2/v$ is larger or smaller than $\lambda_1/v=0.4$.

\section{Material realizations \label{sec:DFT}}

In this final section, we propose new material realizations of noncentrosymmetric Dirac semimetals. In particular, we propose two materials, Bi$_{2}$PdO$_{4}$ and the LiZnSb$_{x}$Bi$_{1-x}$ alloy, which have fourfold and a sixfold rotation axis, respectively.

\subsection{Computational details of first principles calculations}
To predict real materials hosting noncentrosymmetric topological Dirac semimetals, we carried out the structural optimization and electronic structure calculations within the framework of density functional theory (DFT)~\cite{Tong66p1, Hohenberg64pB864}, using the Vienna {\it ab initio} simulation package (\textsc{VASP}) \cite{Kresse96p11169} based on the projector augmented wave (PAW) method \cite{Blochl94p17953}. The exchange-correlation interaction was treated within the generalized gradient approximation (GGA) parametrized by Perdew, Burke, and Ernzerhof (PBE)~\cite{Perdew96p3865}. The energy cutoff of 500\,eV was set in all the calculations and the spin-orbit coupling interactions are considered in the calculations of electronic structures. For the structural optimization, the lattice parameters and all the atoms are relaxed until the Hellmann-Feynman forces on all atoms are less than 0.005\,eV/\AA. A $k$-point mesh was used for Brillouin zone integration and we used $7 \times 7 \times 7$ and $15 \times 15 \times 9$ Monkhorst-Pack grids for Bi$_{2}$PdO$_{4}$ and LiZnSb$_{x}$Bi$_{1-x}$, respectively. We calculated dynamic properties of Bi$_{2}$PdO$_{4}$ by the finite displacement method \cite{Parlinski97p4063}, as implemented in the Phonopy code \cite{togo15p1}. In order to check the bands inversion in Bi$_{2}$PdO$_{4}$ and LiZnBi, we also carry out the band structures calculations by nonlocal HeydScuseria-Ernzerhof (HSE06) hybrid functional method \cite{heyd03p8207}.\par

\subsection{Noncentrosymmetric Dirac semimetal \texorpdfstring{Bi$_{2}$PdO$_{4}$}{BiPdO}}

\begin{figure}[pt]
    \centering
    \includegraphics[width=\columnwidth]{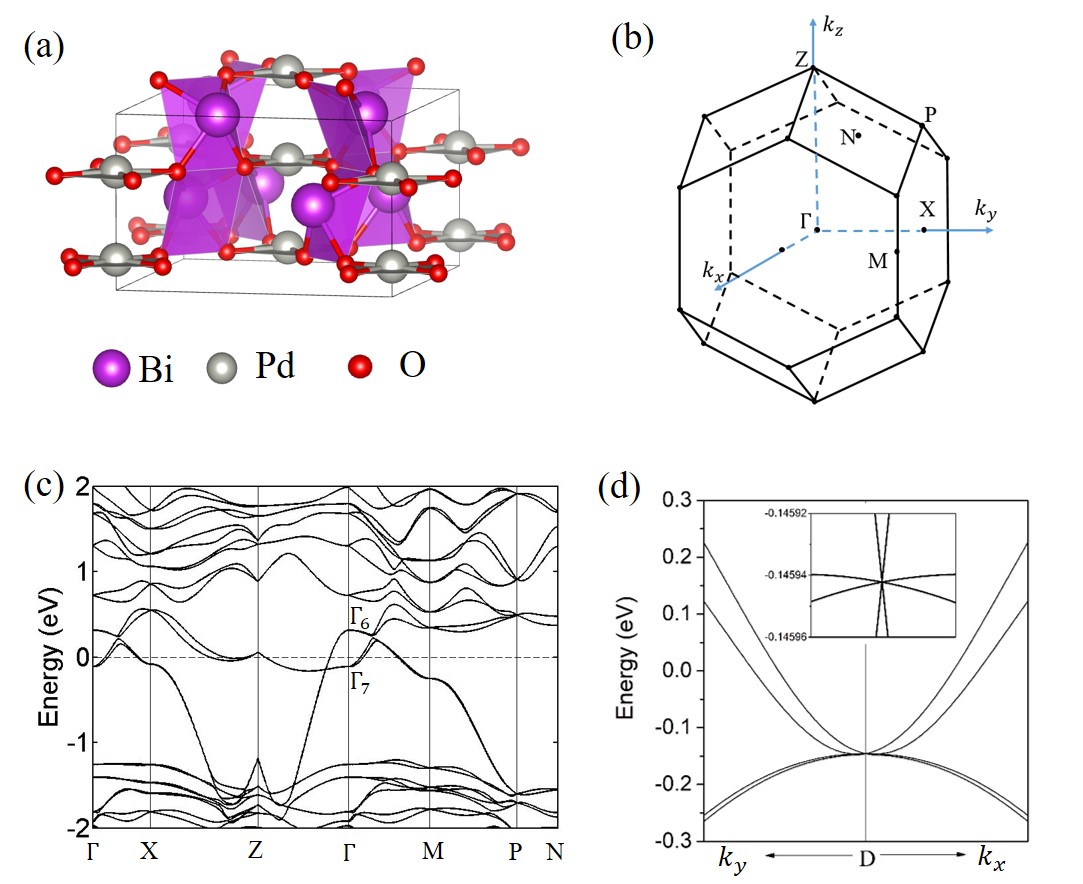}
    \caption{(a) Crystal structure of Bi$_{2}$PdO$_{4}$. The purple, gray, and red balls denote Bi, Pd, and O atoms, respectively. (b) The Brillouin zone with high symmetry points of BiPd$_{2}$O$_{4}$. (c) The band structure of Bi$_{2}$PdO$_{4}$ along high symmetry lines. (d) The band dispersions near the Dirac point D along the $k_{x}$ and $k_{y}$ directions. The insert figure is the zoom-in view of band dispersion near the Dirac point.}
    \label{fig:fig_4}
\end{figure}

The crystal structure of Bi$_{2}$PdO$_{4}$ is shown in Fig.~\ref{fig:fig_4}(a). It crystallizes in a tetragonal crystal structure with space group $I4cm$ $(C_{4v}^{10})$ and shows isolated coplanar oxygen polyhedra around Pd$^{2+}$. In experiment, the single crystal Bi$_{2}$PdO$_{4}$ has been synthesized by a mixture of the oxides PdO and Bi$_{2}$O$_{3}$ at high temperature \cite{Arpe76p1708}. Note that the Bi$_{2}$PdO$_{4}$ has a centrosymmetric allotrope with space group $P4/ncc$, which is a semiconductor according to first-principles calculations \cite{He17p2529}. The calculated phonon dispersion of noncentrosymmetric Bi$_{2}$PdO$_{4}$, presented in Appendix \ref{app:phonon}, shows no imaginary frequency and thus implies that noncentrosymmetric Bi$_{2}$PdO$_{4}$ is dynamically stable.

Figure \ref{fig:fig_4}(c) shows the PBE band structure of Bi$_{2}$PdO$_{4}$ in the presence of spin-orbit interaction. The band inversion between conduction and valence bands occur near $\Gamma$ point, giving rise to electron pockets along the $\Gamma$-Z line and hole pockets along the $\Gamma$-X and $\Gamma$-M lines. These features indicate that Bi$_2$PdO$_{4}$ is a semimetal. Due to the noncentrosymmetric structure and the spin-orbit interaction in Bi$_{2}$PdO$_{4}$, the energy bands are non-degenerate except for the time-reversal invariant points and lines of high symmetry. In particular, the bands along the $\Gamma$-Z line, which is the $C_{4z}$-invariant axis with little group $C_{4v}$, remain doubly degenerate. The conduction and valence bands which are inverted at $\Gamma$ have symmetry character $\Gamma_6$ and $\Gamma_7$, which corresponds to $j_z=\pm \frac12$ and $j_z=\pm \frac32$ states, respectively. The $\Gamma_7$ have predominantly Bi-$p$ character, whereas the $\Gamma_6$ are predominantly made up of Pd-$d$ states. As a result of this band inversion, the band crossing on the $\Gamma$-Z line is protected by symmetry, and this crossing realizes a noncentrosymmetric Dirac point of the kind discussed in Secs.~\ref{sec:ham} and \ref{sec:analysis}. Note, however, that in this case, the Dirac point is of type-II\cite{Soluyanov15p495,Huang16p121117,Chang17p026404}. Figure \ref{fig:fig_4}(d) presents the band dispersions near Dirac points along the $k_{x}$ and $k_{y}$ directions. Each band is singly degenerate, which is consistent with discussion of Secs.~\ref{sec:ham}. Moreover, the Dirac cone is overtilted along the $\Gamma$-A direction and shows the electronic and hole pockets coexist near Dirac point, which are indeed  the characteristic features of the type-II Dirac fermions. As we have known, the PBE functional underestimate the band gap of semiconductor and overestimate the band inversion gap of topological materials. Then we have performed the HSE06 calculation to check the band structure of Bi$_2$PdO$_{4}$. It turns out that Bi$_2$PdO$_{4}$ is an indirect semiconductor with a band gap of 0.58 eV. To realize the band inversion in Bi$_2$PdO$_{4}$, we study the pressure effect to the electronic structures and find that the Bi$_2$PdO$_{4}$ is a type-II noncentrosymmetric Dirac semimetal under 20 Gpa pressure. The HSE06 band structures of Bi$_2$PdO$_{4}$ under different pressures are presented in Appendix \ref{app:hse06}

\subsection{Tunable Dirac and Weyl Fermions in \texorpdfstring{LiZnSb$_{x}$Bi$_{1-x}$ alloy}{LiZnSbBi}}

\begin{figure}[pt]
    \centering
    \includegraphics[width=8.6cm]{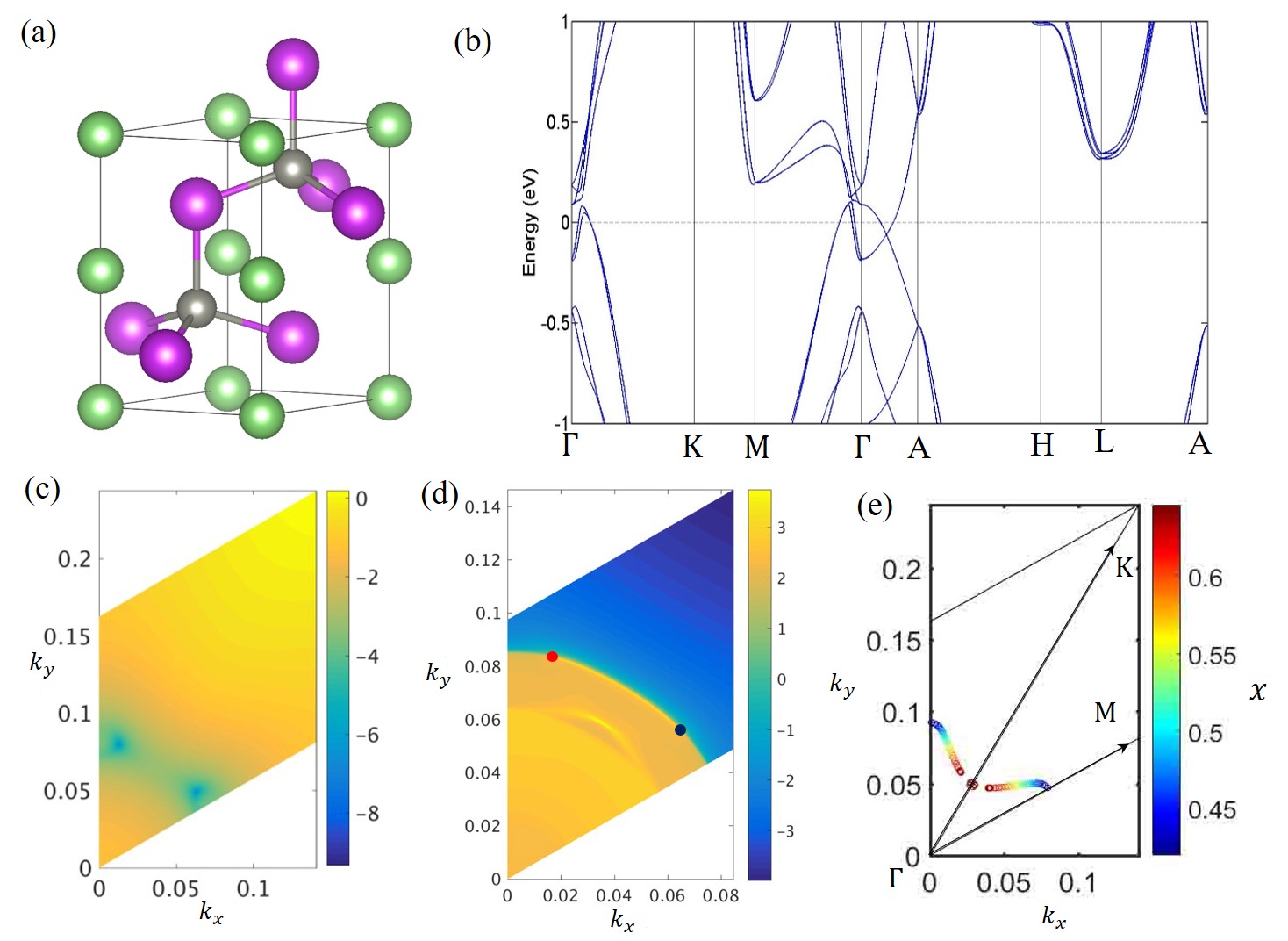}
    \caption{(a) Crystal structure of hexagonal polar LiZnSb$_{x}$Bi$_{1-x}$ alloy. The green, gray, and purple balls denote Li, Zn, and Sb or Bi atoms. (b) The band structure of LiZnSb$_{0.5}$Bi$_{0.5}$ alloy. (c) The energy difference between conduction and valence bands in the $k_{z}=0$ plane of LiZnSb$_{0.5}$Bi$_{0.5}$ alloy. Here the colormap shows logarithmic scale of energy difference log$(E_{cb}-E_{vb})$. The surface states of (d) top and (e) bottom surfaces of LiZnSb$_{x}$Bi$_{1-x}$ alloy. The red and blue dots indicate the projected Weyl points on the surfaces. (f) The positions of Weyl points in the $k_{z}=0$ plane with respect to the alloy concentration $x$ in LiZnSb$_{x}$Bi$_{1-x}$. The color values denote the alloy concentration $x$.}
   \label{fig:fig_5}
\end{figure}

Both LiZnBi and LiZnSb crystallize in the $ABC$ hexagonal polar space group $P6_{3}mc$. The crystal structure of LiZnBi(Sb), known as the stuffed wurtzite lattice, is shown in Fig.~\ref{fig:fig_5}. The Zn and Bi (Sb) atoms form the wurtzite structure and the Li atoms occupy the interstitial sites of the wurtzite lattice. Both LiZnBi and LiZnSb have been experimentally synthesized.\cite{schroeder75p978,tiburtius78p35} LiZnBi has been identified as a Dirac semimetal by first principles calculations in previous work\cite{Cao17p115203} and is thus predicted to realize a noncentrosymmetric Dirac semimetal with $C_{6v}$ point group. 

Moreover, a number of other hexagonal ABC crystals have been predicted to realize Dirac semimetals, including a SrHgPb family\cite{Gao18p106404}, CaAgBi\cite{Chen17p044201}, and LiGaGe\cite{Hu19p195154}. Interestingly, the SrHgPb family of materials have been proposed to host coexisting Dirac and Weyl points, the latter of which are tunable by the HgPb buckling \cite{Gao18p106404}. The regulation of HgPb buckling is however difficult to control in experiment. By contrast, a feasible route to controlling the existence and properties of coexisting Dirac and Weyl points is alloy engineering, as has been pointed by previous theoretical predictions \cite{Huang18p136403,Fang20p125202} and demonstrated in experiments \cite{Zhang11p1,Sato11p840}. Here we propose alloy engineering of LiZnBi and LiZnSb.

In contrast to LiZnBi, LiZnSb is a semiconductor with the band gap of 0.42 eV. The band structure and Wannier charge center (WCC) of LiZnSb are presented in Appendix \ref{app:band}, showing that LiZnSb is a topologically trivial semiconductor. So one should expect that the electronic structures and topological transition are tunable by changing the alloy concentration in LiZnSb$_{x}$Bi$_{1-x}$. To confirm this idea, we calculate the electronic structures of LiZnSb$_{x}$Bi$_{1-x}$ alloy using the virtual crystal approximation method \cite{Porod83p2587}. The effective Hamiltonian of LiZnSb and LiZnBi are obtained using Maximum Localized Wannier Function (MLWF) as implemented in Wannier90 package \cite{Mostofi08p685}. Then the effective Hamiltonian of LiZnSb$_{x}$Bi$_{1-x}$ alloy is linearly interpolated between LiZnSb and LiZnBi ones. The band structure of LiZnSb$_{0.5}$Bi$_{0.5}$ is calculated and shown in Fig.~\ref{fig:fig_5}(b). It turns out the LiZnSb$_{0.5}$Bi$_{0.5}$ is a noncentrosymmetric topological Dirac semimetal. The conduction and valence bands remain inverted and the Dirac points along the $\Gamma$-X line are symmetry-protected by the $C_{6v}$ points group. Compared with the pristine undoped Dirac semimetal LiZnBi, the energy level of Dirac points in LiZnSb$_{0.5}$Bi$_{0.5}$ is exactly tuned closer to the Fermi level. This is a great advantage in the experiment to measure the transport properties of noncentrosymmetric topological Dirac semimetals. \par

More interestingly, we found that the Weyl points also appear in the LiZnSb$_{0.5}$Bi$_{0.5}$ alloy. To prove the existence of Weyl points, the energy difference between conduction and valence bands in the $k_z$ are calculated and shown in the Fig.~\ref{fig:fig_5}(c). It turns out that there are six pairs of Weyl points in the $k_{z}$ plane. These twelve Weyl points are related by $C_{6v}$ and time-reversal symmetries. The positions of one independent pair of Weyl points are identified as ($\pm 0.018$, 0.085, 0) in fractional coordinates. To further confirm the Weyl semimetal phase of LiZnSb$_{0.5}$Bi$_{0.5}$ alloy, the topological Fermi arc surface states of the top and bottom surfaces are calculated and presented in Fig.~\ref{fig:fig_5}(d)--(e). Since the topological properties of LiZnSb and LiZnBi show significant difference, one may continually change the alloy concentration in the LiZnSb$_{1-x}$Bi$_{x}$ to explore its topological properties by first principles calculations or in the experiments. By calculating the electronic structures of LiZnSb$_{1-x}$Bi$_{x}$ with the continuous change of $x$ value from 0 to 1, we find that the Dirac points appear when the value is larger then 0.38 and the Weyl points are generated on the $\Gamma$-M line when $x$ value equals 0.41. The trace of Weyl points with alloy concentration $x$ value is shown in Fig.~\ref{fig:fig_5}(d). One can see that each pair of Weyl points move from $\Gamma$-M line to $\Gamma$-K line with the increase of the value of $x$ and annihilate at $\Gamma$-K line. So the topological phases of Dirac and Weyl points coexsit when the value of $x$ is between 0.41 and 0.65. The first-principles calculations are consistent with our $k \cdot p$ model as discussed in Sec. \ref{ssec:derivation}.\par

\section{Discussion and conclusion \label{sec:conclusion}}

In this paper, we have developed a general study of topological Dirac semimetals in noncentrosymmetric crystals geared towards material realizations and prediction. We have identified the crystallographic point group symmetry requirements for stable fourfold degenerate band crossings on the rotation axis induced by a band inversion at $\Gamma$, and have identified all distinct types of band inversions based on the band angular momentum quantum numbers. For each type, we derived model Hamiltonians describing the inverted bands, using formulation which makes the physical nature of the allowed couplings transparent, and thus allows for a systematic study of the distinct semimetallic phases which can generically occur as a function of model parameters. 

We have established the generic phase diagram of noncentrosymmetric topological Dirac semimetals by studying the model Hamiltonians in two different ways: close to the Dirac point band crossings and on the mirror planes. Our analysis shows that, depending on the strength of intra-band spin-orbit coupling allowed by inversion symmetry breaking, the Dirac points can change character and become attached to Weyl line nodes on the mirror planes. Weyl line nodes are generically present in the phase diagram of noncentrosymmetric Dirac semimetals, enabled by intra-band spin-orbit coupling. For one type of band inversion, we have shown that Weyl line nodes connecting to the Dirac points always occur. We further determine that Weyl point nodes coexisting with the Dirac points are generically present in the phase diagram. The presence of Weyl points depends on the precise strength of the coupling between the inverted bands. 

Two material candidates, Bi$_{2}$PdO$_{4}$ under 20 Gpa pressure and the LiZnSb$_{x}$Bi$_{1-x}$ alloy, are proposed as new realizations of noncentrosymmetric Dirac semimetals. Using first-principles materials prediction, we find that Bi$_{2}$PdO$_{4}$ with point group $C_{4v}$ hosts the type-II Dirac Fermions on the $C_{4z}$ rotational axis, and further find that the alloy LiZnSb$_{x}$Bi$_{1-x}$ with point group $C_{6v}$ can realize doping-tunable Weyl points on the $k_z=0$ plane of momentum space. 

We conclude by noting that model Hamiltonians derived in this work provide a fruitful basis for further study of noncentrosymmetric Dirac semimetals, in particular with regard to properties related to inversion symmetry breaking. For instance, very recently a large nonlinear optical response \cite{Wu17p350,Ma19p476} and quantum nonlinear Hall effect \cite{Sodemann15p216806,Ma19p337} have been predicted for Weyl semimetals which preserve time-reversal symmetry but break inversion symmetry. It will be very interesting to explore similar properties and effects in noncentrosymmetric topological Dirac semimetals \cite{Ahn20p041041}, and our analysis provides a natural framework for doing so. In this regard, it is worth mentioning that our derivation of $k\cdot p$-type model Hamiltonians can be directly and straightforwardly extended to obtain full lattice models (i.e., tight-binding models) for the relevant low-energy bands. 

\begin{acknowledgments}
We acknowledge the supports from the National Natural Science Foundation (Grant No.11925408), the Ministry of Science and Technology of China (Grants No.2016YFA0300600 and 2018YFA0305700), the Chinese Academy of Sciences (Grant No.XDB33000000), the K. C. Wong Education Foundation (GJTD-2018-01), the Beijing Natural Science Foundation (Z180008), and the Beijing Municipal Science and Technology Commission (Z191100007219013). H.G. acknowledges support from the National Postdoctoral Program for Innovative Talents (No.BX20190361) and Guangdong Basic and Applied Basic Research Foundation (No. 2019A1515110965). The computational resource is provided by the Platform for Data-Driven Computational Materials Discovery in Songshan Lake material Laboratory.
\end{acknowledgments}

\appendix

\section{Single-band Hamiltonian \texorpdfstring{$h_\bk$}{hk} \label{app:h_k}}

In this appendix we describe the derivation of the single-band Hamiltonians $h_\bk$ given the symmetry type of the band (i.e., the value of $j_z$). In fact, we only need to find $\bb_\bk$. 

First, consider a band of $j_z=\pm\frac12$ states. Rotational symmetry (either fourfold $C_{4z}$ or sixfold $C_{6z}$) mandates the form
\be
b^x_\bk-i b^y_\bk = \lambda k_-.
\ee
From \eqref{h-mirror} it follows that $b^x_\bk +i b^y_\bk =- \lambda k_+$, but since also $b^x_\bk +i b^y_\bk =\lambda^* k_+$, one must have $\lambda^*=-\lambda$, i.e., $\lambda$ is purely imaginary. We thus find that 
\be
\bb_\bk \cdot \makebf{\sigma} = \bar \lambda(k_x\sigma_y - k_y\sigma_x),
\ee
with $ \bar \lambda$ real.

In the case of $j_z=\pm\frac32$ states rotational symmetry matters. In particular, $C_{4z}$ symmetry mandates the form
\be
b^x_\bk-i b^y_\bk = \lambda k_+,
\ee
whereas $C_{6z}$ symmetry mandates the form
\be
b^x_\bk-i b^y_\bk = \lambda_1 k^3_- + \lambda_2 k^3_+.
\ee
As before, mirror symmetry requires that $\lambda^*=-\lambda$, and further that $\lambda_{1,2}^*=-\lambda_{1,2}$. In the case of fourfold rotation symmetry we simply have $\bb_\bk \cdot \makebf{\sigma} = \bar \lambda(k_x\sigma_y + k_y\sigma_x)$, with $ \bar \lambda$ real. Instead, in the case of sixfold symmetry we have
\be
\bb_\bk \cdot \makebf{\sigma} =-i  \bar \lambda(k^3_+ -k^3_-)\sigma_x - \bar \lambda'(k^3_+ +k^3_-)\sigma_y,
\ee
with $ \bar \lambda,\bar \lambda' $ real. As a result, the breaking of inversion symmetry does not lead to a linear splitting of Kramers pairs away from time-reversal invariant momenta. 

Finally, consider a band of $j_z=\pm\frac52$ states, where only the case of $C_{6x}$ symmetry is relevant. For sixfold rotations we have 
\be
b^x_\bk-i b^y_\bk = \lambda k_+,
\ee
and from \eqref{h-mirror} we again find that $\lambda^*=-\lambda$.

\section{Band inversion of \texorpdfstring{$j_z=\pm\frac12$}{jz12}  and \texorpdfstring{$j_z=\pm\frac32$}{jz52} bands \label{app:1/2-3/2}}

We consider a band inversion of $j_z=\pm\frac12$  and $j_z=\pm\frac32$ bands and determine the form of $\Delta_\bk$. Rotation symmetry (either fourfold or sixfold) mandates the general form
\be
\Delta_\bk = \begin{pmatrix} \gamma_1k_+ & \gamma_3 k^2_-+\gamma'_3 k^2_+ \\  \gamma_4 k^2_++\gamma'_4 k^2_- & \gamma_2 k_-\end{pmatrix}
\ee
with coefficients $\gamma_i=\gamma_i(k_z)$ which may still  be functions of $k_z$. Note that in the case of sixfold $C_{6z}$ symmetry we must have $\gamma'_3=\gamma'_4=0$. Connecting this with the expansion of \eqref{D-expand} we find that $\delta_\bk$ and $d^z_\bk$ are given by
\be
\delta_\bk = \frac12(\gamma_1 k_+ + \gamma_2 k_-), \quad d^z_\bk = \frac12(\gamma_1 k_+ - \gamma_2 k_-).
\ee
Time reversal invariance, as given by Eq.~\eqref{D-time}, leads to the condition
\be
\gamma^*_2(k_z) = -\gamma_1(-k_z).
\ee
Invariance under mirror symmetry $M_x$ implies 
\be
\gamma_2(k_z) = \gamma_1(k_z) ,
\ee
and combining these two conditions we obtain the constraint
\be
\gamma^*_1(k_z)=-\gamma_1(-k_z).
\ee
We conclude that when expanding $\gamma_1(k_z)$ in $k_z$, the even terms have purely imaginary coefficients, whereas the odd terms are purely real. 

For the off-diagonal terms of $\Delta_\bk$ we have
\be
d^x_\bk-id^y_\bk = \gamma_3 k^2_-+\gamma'_3 k^2_+.
\ee
Time-reversal symmetry enforces the constraint
\be
-(d^x_\bk+id^y_\bk)^* = d^x_{-\bk}-id^y_{-\bk},
\ee
which leads to the condition
\be
\gamma^*_4(k_z) = -\gamma_3(-k_z), \quad {\gamma'_4}^*(k_z) = -\gamma'_3(-k_z).
\ee
Mirror symmetry implies the conditions
\be
\gamma_4 = -\gamma_3, \quad \gamma'_4 = -\gamma'_3,
\ee
and these may be combined with the time-reversal condition to obtain
\be
\gamma^*_3(k_z) = \gamma_3(-k_z),
\ee
and the same relation for $\gamma'_3$. Hence, the even terms in $k_z$ have purely real coefficients, whereas the odd terms are purely imaginary.

\section{Band inversion of \texorpdfstring{$j_z=\pm\frac12$}{jz12}  and \texorpdfstring{$j_z=\pm\frac52$}{jz52} bands \label{app:1/2-5/2}}

Consider next a band inversion of $j_z=\pm\frac12$  and $j_z=\pm\frac52$. In this case, we only need to consider sixfold rotationally symmetric systems, since $j_z=\pm\frac52$ doublets fall in the same symmetry class as $\pm\frac32$ states in $C_4$ systems (i.e., they have the same representation). Sixfold rotation symmetry mandates the general form
\be
\Delta_\bk = \begin{pmatrix} \gamma_1k^2_+ & \gamma_3 k^3_-+\gamma'_3 k^3_+ \\  \gamma_4 k^3_++\gamma'_4 k^3_- & \gamma_2 k^2_-\end{pmatrix}
\ee
where the coefficients $\gamma_i=\gamma_i(k_z)$ are again functions of $k_z$. In terms of the expansion of \eqref{D-expand} we find that $\delta_\bk$ and $d^z_\bk$ are given by
\be
\delta_\bk = \frac12(\gamma_1 k^2_+ + \gamma_2 k^2_-), \quad d^z_\bk = \frac12(\gamma_1 k^2_+ - \gamma_2 k^2_-).
\ee
Time-reversal invariance, as expressed in Eq.~\eqref{D-time}, leads to the condition
\be
\gamma^*_2(k_z) = \gamma_1(-k_z).
\ee
Invariance under mirror symmetry $M_x$ implies the condition
\be
\gamma_2(k_z) = \gamma_1(k_z) ,
\ee
and combining these two conditions we obtain the single condition
\be
\gamma^*_1(k_z)=\gamma_1(-k_z).
\ee
We conclude that when expanding $\gamma_1(k_z)$ in $k_z$, the even terms have purely real coefficients, whereas the odd terms are purely imaginary. 

For the off-diagonal terms of $\Delta_\bk$ we have
\be
d^x_\bk-id^y_\bk = \gamma_3 k^3_-+\gamma'_3 k^3_+.
\ee
From time-reversal symmetry we find
\be
\gamma^*_4(k_z) = \gamma_3(-k_z), \quad {\gamma'_4}^*(k_z) = \gamma'_3(-k_z)
\ee
Mirror symmetry implies the conditions
\be
\gamma_4 = -\gamma_3, \quad \gamma'_4 = -\gamma'_3,
\ee
and these may be combined with the time-reversal condition to obtain
\be
\gamma^*_3(k_z) = -\gamma_3(-k_z),
\ee
and the same relation for $\gamma'_3$. Hence, the even terms in $k_z$ have purely imaginary coefficients, whereas the odd terms are purely real

\section{Inversion symmetric limits \label{app:inversion}}

In principle, the inversion symmetric limit of our model Hamiltonians is ambiguous, since the relative parity of the two bands is ill-defined. By treating the two possible cases separately, here will show that the two different ways of taking the inversion symmetric limit can be considered equivalent. 

{\it Band inversion of $j_z=\pm\frac12$ and $j_z=\pm\frac32$ bands.} First, consider the band inversion of $j_z=\pm\frac12$ and $j_z=\pm\frac32$ bands, for which the coupling matrix $\Delta_\bk$ is given by Eq.~\eqref{eq:Delta-1/2-3/2}. If we assume that, in the inversion symmetric limit, the two bands have equal parity, then the Hamiltonian $\mathcal  H^\pm_\bq$ near the two Dirac points at $\pm \bK_0$ is given by
\be
\mathcal  H^\pm_\bq= \pm ( vq_z \tau_z + \tilde\alpha_2q_x\tau_x -  \tilde\alpha_2q_y\tau_y\sigma_z),  \label{app:H_equal_P}
\ee
where $\tau_z=\pm1$ labels the conduction and valence bands, and $\tilde\alpha_2 =\alpha_2 K_0$. The velocity $v$ is defined as $v=K_0/m_z$. To isolate the two Weyl points contained in this Dirac point, we seek a chiral operator $\Gamma$ which commutes with $\tau_z$, $\tau_x$, and $\tau_y\sigma_z$, and satisfies $\Gamma^2=1$. The operator with the desired properties is given by $\Gamma = -\sigma_z$. Writing the full noncentrosymmetric Hamiltonian near each Dirac point $\pm\bK_0$ in a basis of chiral eigenstates of $\Gamma$ brings the Hamiltonian into the form of Eq.~\eqref{eq:H_q}, with $A^\pm_\bq= q_iA^\pm_{ij}\tau_j $. (Note that here, in the case of $A^\pm_\bq$, $\pm$ refers to the chirality of the Weyl points forming a Dirac point.) At the Dirac point located at $+\bK_0$, $A^\pm_\bq$ is given by Eq.~\eqref{eq:A_pm_1/2-3/2}, where as at $+\bK_0$ one has
\be
A^\pm = \begin{pmatrix} -\tilde \alpha_2 & -\alpha_1 & 0 \\ \pm \alpha_1  &\mp \tilde \alpha_2& 0\\ 0 &0& -v \end{pmatrix}.
\ee

Now consider the other possible inversion symmetric limit, in which the two bands have opposite parity. In that case, the inversion symmetric Hamiltonian at $\pm \bK_0$ is given by
\be
\mathcal  H^\pm_\bq =\pm  vq_z \tau_z -\alpha_1q_x\tau_y -\alpha_1q_y\tau_x\sigma_z \label{app:H_opposite_P}
\ee
The chiral operator $\Gamma$ can be taken as before, $\Gamma=-\sigma_z$, and this leads to the same expressions for $A^\pm_\bq$ and $B_\bq$. We thus conclude that, in this sense, the two inversion symmetric limits are equivalent.

{\it Band inversion of $j_z=\pm\frac12$ and $j_z=\pm\frac52$ bands.} In the case of the second type of band inversion, between $j_z=\pm\frac12$ and $j_z=\pm\frac52$ bands, taking the inversion symmetric limit(s) introduces an additional subtlety. Consider first the case of equal parity bands. In this case, the Hamiltonian near the two Dirac points at $\pm \bK_0$ is
\be
\mathcal  H^\pm_\bq= \pm  vq_z \tau_z + \alpha_1(q^2_x-q^2_y)\tau_x -  2\alpha_1q_xq_y\tau_y\sigma_z,  \label{app:H_equal_P_2}
\ee
At $+ \bK_0$ the chiral operator can again be chosen as $\Gamma=-\sigma_z$, and this leads to the matrices $A^\pm_\bq$ (at $+ \bK_0$) given by 
\be
A^\pm_\bq = vq_z\tau_z  +   \alpha_1 (q^2_\mp\tau_+ +    q^2_\pm \tau_- ). \label{app:A_pm_1/2-5/2_a_1}
\ee
Here we have introduced $\tau_\pm  = (\tau_x \pm i \tau_y)/2$ and $q_\pm = q_x \pm i q_y$. Importantly, the matrices $A^\pm_\bq $ describe a topological band crossing with monopole charge $C=\pm 2$, which is a result of the quadratic dispersion in the plane. This should be contrasted with the case of linear dispersion.  

In the converse case, when the parity eigenvalues are opposite, the Hamiltonian near $\pm \bK_0$ takes the form
\be
\mathcal  H^\pm_\bq= \pm  [ vq_z \tau_z -\tilde \alpha_2(q^2_x-q^2_y)\tau_y -  2\tilde \alpha_2 q_xq_y\tau_x\sigma_z ],  \label{app:H_opposite_P_2}
\ee
with $\tilde \alpha_2 = \alpha_2 K_0$. The chirality operator is taken as $\Gamma=-\sigma_z$, as before, and we find the matrices $A^\pm_\bq$ as
\be
A^\pm_\bq = vq_z\tau_z  +   \tilde \alpha_2 i (q^2_\mp\tau_+ - q^2_\pm \tau_- ). \label{app:A_pm_1/2-5/2_b}
\ee

Starting from either of these two limits one may activate the inversion symmetry breaking terms and arrive at the full form of $A^\pm_\bq$ given by
\be
A^\pm_\bq = vq_z\tau_z  +    \bar \alpha q^2_\mp\tau_+ +    \bar \alpha^* q^2_\pm \tau_- , \label{app:A_pm_1/2-5/2_a_2}
\ee
with $ \bar \alpha = \alpha_1 +i \alpha_2 K_0$.

\section{Diagonalization of \texorpdfstring{$H_\bk$}{Hk} \label{app:diag}}

Taking the Hamiltonian of Eq.~\eqref{eq:Hk} and setting $\lambda_v=\lambda_c=0$ one obtains
\be
H_\bk = \begin{pmatrix} \varepsilon_\bk & \Delta_\bk \\ \Delta^\dagger_\bk  &  -\varepsilon_\bk  \end{pmatrix},
\ee
which can be straightforwardly diagonalized by using well-known properties of non-unitary pairing states. For the matrix product $\Delta_\bk \Delta^\dagger_\bk$ we find
\be
\Delta_\bk \Delta^\dagger_\bk= |\delta_\bk|^2 +| \bd_\bk |^2 +(\delta^*_\bk\bd_\bk+\delta_\bk\bd^*_\bk+i\bd_\bk\times \bd^*_\bk)\cdot \bsigma
\ee
and leads to an equation for the energies $E_\bk$ given by
\be
E^2_\bk = \varepsilon^2_\bk +|\delta_\bk|^2 +| \bd_\bk |^2 \pm F_\bk
\ee
with $F_\bk$ defined as
\begin{multline}
F_\bk = \left[( \delta^*_\bk)^2\bd^2_\bk+ \delta_\bk^2(\bd^*_\bk)^2 + 2|\delta_\bk|^2 | \bd_\bk |^2 \right. \\
\left. - |\bd^2_\bk|^2 + |\bd_\bk|^4\right]^{1/2}.
\end{multline}

In the present context, these solutions can be further simplified when $k_z=0$. In this case we find that $\delta_\bk$ and the components of $\bd_\bk$ are either purely real or purely imaginary. In particular, either $\delta_\bk,d^z_\bk $ are purely real and $d^x_\bk, d^y_\bk$ are purely, or vice versa. In either case $F_\bk$ reduces to
\be
F_\bk = 2\sqrt{(|\delta_\bk|^2+|d^z_\bk|^2)(|d^x_\bk|^2+|d^y_\bk|^2)},
\ee
which may be used to rewrite $E^2_\bk$ as
\be
E^2_\bk  = \varepsilon^2_\bk + \left( \sqrt{|\delta_\bk|^2+|d^z_\bk|^2}\pm  \sqrt{|d^x_\bk|^2+|d^y_\bk|^2}\right)^2.
\ee
By substituting the appropriate expressions for $\delta_\bk$ and $\bd_\bk$ we obtain Eqs.~\eqref{eq:Ek_C4} and \eqref{eq:Ek_C6} of the main text.

\begin{figure}[t]
    \centering
    \includegraphics[width=8.6cm]{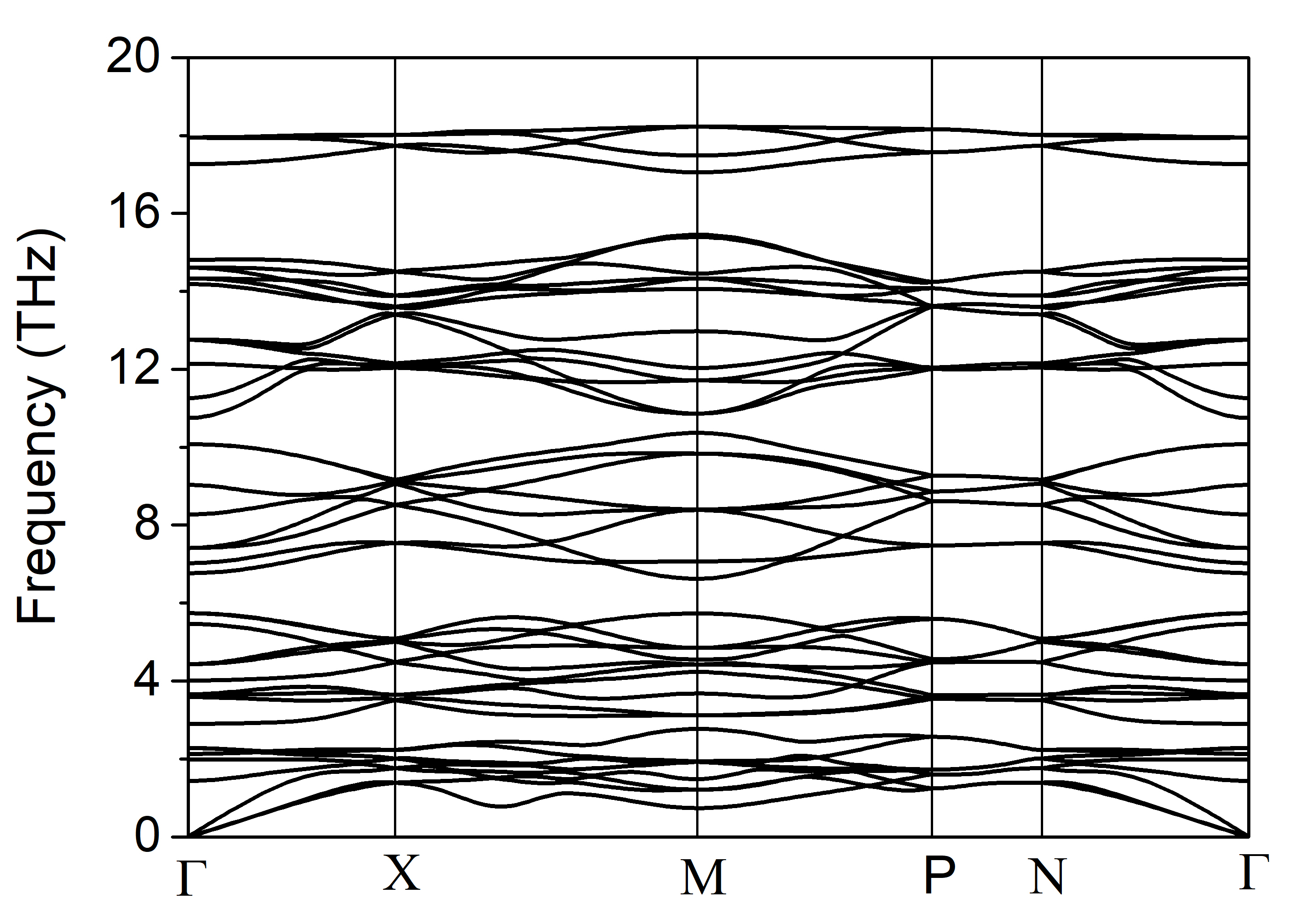}
    \caption{The phonon dispersion of Bi$_2$PdO$_4$.}
   \label{fig:fig_8}
\end{figure}

\section{Phonon dispersion of \texorpdfstring{Bi$_2$PdO$_4$}{BiPdO} \label{app:phonon}}
To illustrate the dynamic properties of Bi$_2$PdO$_4$, the phonon dispersion was calculated using the finite difference method with a $2 \times 2 \times 2$ supercell and shown in the Fig.~\ref{fig:fig_8}. No imaginary frequency is observed in the dispersion. It turns out that Bi$_2$PdO$_4$ is dynamic stable.

\section{HSE06 band structures of \texorpdfstring{Bi$_2$PdO$_4$}{BiPdO} under pressure \label{app:hse06}}
\begin{figure}[t]
    \centering
    \includegraphics[width=8.6cm]{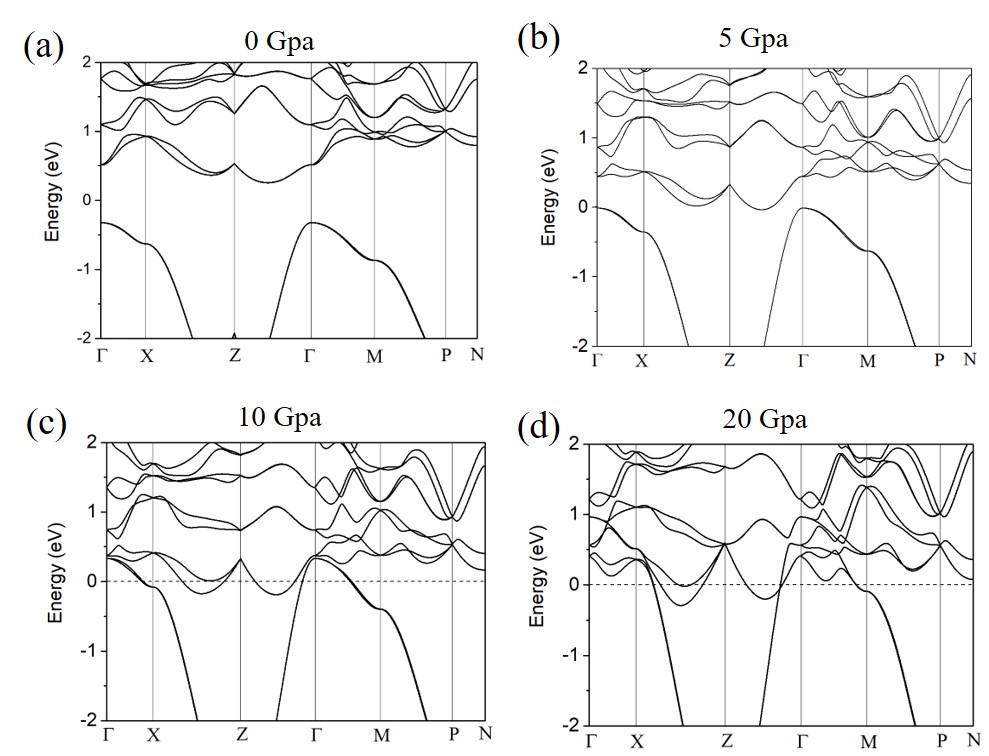}
    \caption{HSE06 band structures of Bi$_2$PdO$_4$ with the pressures of (a) 0 Gpa, (b) 5 Gpa, (c) 10 Gpa, and (d) 20 Gpa}
   \label{fig:fig_9}
\end{figure}

In order to study the electronic structures of Bi$_2$PdO$_4$ under different pressures, the band structures with the pressures of 0 Gpa, 5 Gpa, 10 Gpa, and 20 Gpa were calculated by the HSE06 method and presented in Fig.~\ref{fig:fig_9}. We found that Bi$_2$PdO$_4$ is a semiconductor and become a semimetal when the pressure is larger than 5 Gpa. However, there is no band inversion at $\Gamma$ point in Bi$_2$PdO$_4$ when the pressure is 5 Gpa. This is due to that Bi$_2$PdO$_4$ is an indirect semiconductor. When the pressure is continually increased to larger than 10 Gpa, one will observe $\Gamma$ point is close to the band inversion between conduction and valence bands. In the case of 20 Gpa pressure, Bi$_2$PdO$_4$ is a type-II noncentrosymmetric Dirac semimetal. \par

\begin{figure}[b]
    \centering
    \includegraphics[width=8.6cm]{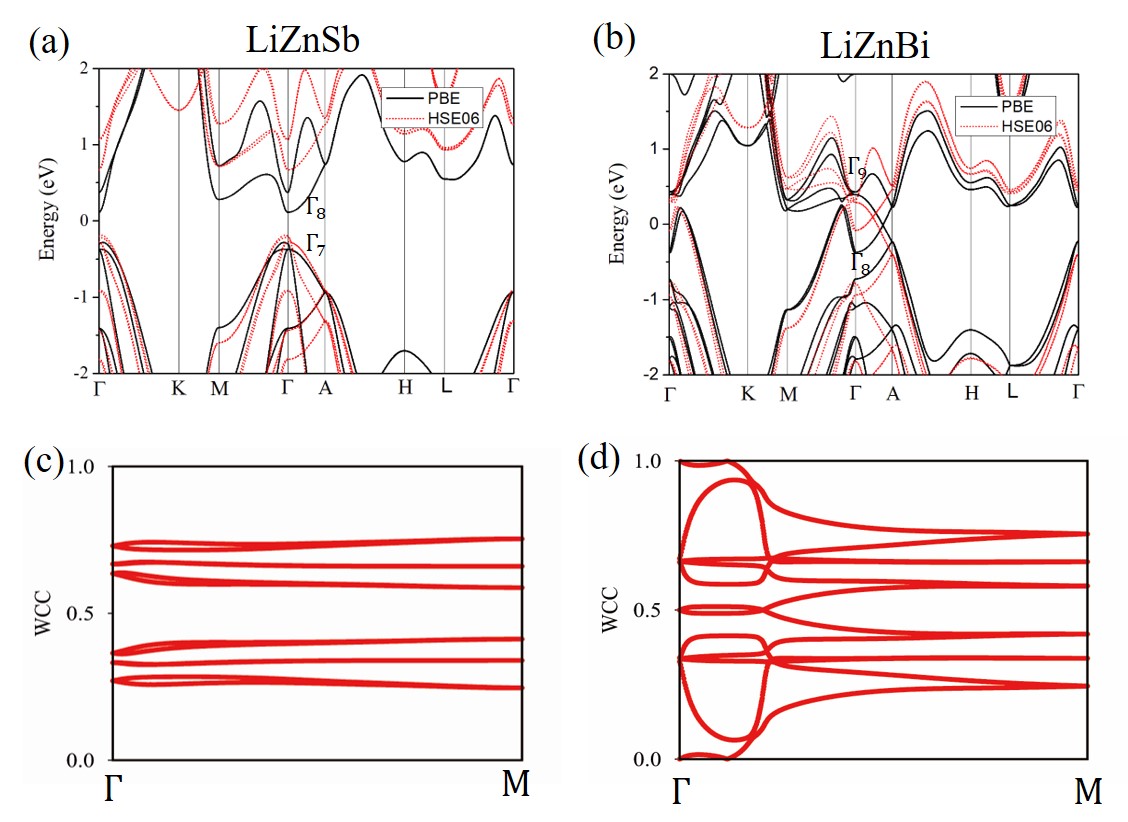}
    \caption{The band structures of (a) LiZnSb and (b) LiZnBi along high symmetry lines. The WCC of (c) LiZnSb (d) LiZnBi in the $k_{z}=0$ plane.}
   \label{fig:fig_10}
\end{figure} 

\section{Band structures and WCC of LiZnSb and LiZnBi \label{app:band}}

To explore the electronic structures and topological properties of LiZnSb and LiZnBi, the bands structures were calculated by PBE and HSE06 methods and shown in Fig.~\ref{fig:fig_10}(a) and (b). The band gaps of LiZnBi are 0.41 eV and 0.87 eV by PBE and HSE06 methods, respectively. The HSE06 band structure of LiZnBi confirms the band inversion with 0.37 eV at $\Gamma$ point. Since the LiZnSb and LiZnBi crystallize with $C_{6v}$ point group which breaks the inversion symmetry. The parity method \cite{Liang07p045302} does not woks here. We employ the Wilson loops method \cite{Soluyanov11p235401,Yu11p075119} to calculate the Wannier Charger center on $k_{z}=0$ plane of LiZnSb and LiZnBi and present in Fig.~\ref{fig:fig_10}(c) and (d). It turns out that $Z_{2}$ invarants on $k_{z}=0$ plane of LiZnSb and LiZnBi are 0 and 1, respectively.

\bibliography{main}

\end{document}